\PassOptionsToPackage{unicode}{hyperref}
\PassOptionsToPackage{hyphens}{url}
\PassOptionsToPackage{dvipsnames,svgnames,x11names}{xcolor}
\documentclass[
]{article}

\usepackage{amsmath,amssymb}
\usepackage{iftex}
\ifPDFTeX
  \usepackage[T1]{fontenc}
  \usepackage[utf8]{inputenc}
  \usepackage{textcomp} % provide euro and other symbols
\else % if luatex or xetex
  \usepackage{unicode-math}
  \defaultfontfeatures{Scale=MatchLowercase}
  \defaultfontfeatures[\rmfamily]{Ligatures=TeX,Scale=1}
\fi
\usepackage{lmodern}
\ifPDFTeX\else  
\fi
\IfFileExists{upquote.sty}{\usepackage{upquote}}{}
\IfFileExists{microtype.sty}{% use microtype if available
  \usepackage[]{microtype}
  \UseMicrotypeSet[protrusion]{basicmath} % disable protrusion for tt fonts
}{}
\makeatletter
\@ifundefined{KOMAClassName}{% if non-KOMA class
  \IfFileExists{parskip.sty}{%
    \usepackage{parskip}
  }{% else
    \setlength{\parindent}{0pt}
    \setlength{\parskip}{6pt plus 2pt minus 1pt}}
}{% if KOMA class
  \KOMAoptions{parskip=half}}
\makeatother
\usepackage{xcolor}
\usepackage[left=20mm,right=20mm]{geometry}
\ifx\paragraph\undefined\else
  \let\oldparagraph\paragraph
  \renewcommand{\paragraph}[1]{\oldparagraph{#1}\mbox{}}
\fi
\ifx\subparagraph\undefined\else
  \let\oldsubparagraph\subparagraph
  \renewcommand{\subparagraph}[1]{\oldsubparagraph{#1}\mbox{}}
\fi

\usepackage{color}
\usepackage{fancyvrb}

\DefineVerbatimEnvironment{Highlighting}{Verbatim}{commandchars=\\\{\}}
\usepackage{framed}
\definecolor{shadecolor}{RGB}{241,243,245}
\newenvironment{Shaded}{\begin{snugshade}}{\end{snugshade}}

\newcommand{\AttributeTok}[1]{\textcolor[rgb]{0.40,0.45,0.13}{#1}}

\newcommand{\CommentTok}[1]{\textcolor[rgb]{0.37,0.37,0.37}{#1}}

\newcommand{\ConstantTok}[1]{\textcolor[rgb]{0.56,0.35,0.01}{#1}}

\newcommand{\DecValTok}[1]{\textcolor[rgb]{0.68,0.00,0.00}{#1}}

\newcommand{\FloatTok}[1]{\textcolor[rgb]{0.68,0.00,0.00}{#1}}
\newcommand{\FunctionTok}[1]{\textcolor[rgb]{0.28,0.35,0.67}{#1}}

\newcommand{\NormalTok}[1]{\textcolor[rgb]{0.00,0.23,0.31}{#1}}

\newcommand{\OtherTok}[1]{\textcolor[rgb]{0.00,0.23,0.31}{#1}}

\newcommand{\SpecialCharTok}[1]{\textcolor[rgb]{0.37,0.37,0.37}{#1}}

\newcommand{\StringTok}[1]{\textcolor[rgb]{0.13,0.47,0.30}{#1}}

\usepackage{longtable,booktabs,array}
\usepackage{calc} % for calculating minipage widths
\usepackage{etoolbox}
\makeatletter
\patchcmd\longtable{\par}{\if@noskipsec\mbox{}\fi\par}{}{}
\makeatother
\IfFileExists{footnotehyper.sty}{\usepackage{footnotehyper}}{\usepackage{footnote}}
\makesavenoteenv{longtable}
\usepackage{graphicx}
\makeatletter
\def\maxwidth{\ifdim\Gin@nat@width>\linewidth\linewidth\else\Gin@nat@width\fi}
\def\maxheight{\ifdim\Gin@nat@height>\textheight\textheight\else\Gin@nat@height\fi}
\makeatother
\setkeys{Gin}{width=\maxwidth,height=\maxheight,keepaspectratio}
\makeatletter
\def\fps@figure{htbp}
\makeatother
\NewDocumentCommand\citeproctext{}{}

\makeatletter
 \let\@cite@ofmt\@firstofone
 \def\@biblabel#1{}
 \def\@cite#1#2{{#1\if@tempswa , #2\fi}}
\makeatother
\newlength{\cslhangindent}
\newlength{\csllabelwidth}
\newenvironment{CSLReferences}[2] % #1 hanging-indent, #2 entry-spacing
 {\begin{list}{}{%
  \setlength{\itemindent}{0pt}
  \setlength{\leftmargin}{0pt}
  \setlength{\parsep}{0pt}
  \ifodd #1
   \setlength{\leftmargin}{\cslhangindent}
   \setlength{\itemindent}{-1\cslhangindent}
  \fi
  \setlength{\itemsep}{#2\baselineskip}}}
 {\end{list}}
\usepackage{calc}

\usepackage{booktabs}
\usepackage{longtable}
\usepackage{array}
\usepackage{multirow}
\usepackage{wrapfig}
\usepackage{float}
\usepackage{colortbl}
\usepackage{pdflscape}
\usepackage{tabu}
\usepackage{threeparttable}
\usepackage{threeparttablex}
\usepackage[normalem]{ulem}
\usepackage{makecell}
\usepackage{xcolor}
\usepackage[noblocks]{authblk}

\makeatletter
\@ifpackageloaded{caption}{}{\usepackage{caption}}
\AtBeginDocument{%
\ifdefined\contentsname
  \renewcommand*\contentsname{Table of contents}
\else
  \newcommand\contentsname{Table of contents}
\fi
\ifdefined\listfigurename
  \renewcommand*\listfigurename{List of Figures}
\else
  \newcommand\listfigurename{List of Figures}
\fi
\ifdefined\listtablename
  \renewcommand*\listtablename{List of Tables}
\else
  \newcommand\listtablename{List of Tables}
\fi
\ifdefined\figurename
  \renewcommand*\figurename{Figure}
\else
  \newcommand\figurename{Figure}
\fi
\ifdefined\tablename
  \renewcommand*\tablename{Table}
\else
  \newcommand\tablename{Table}
\fi
}
\@ifpackageloaded{float}{}{\usepackage{float}}
\floatstyle{ruled}
\@ifundefined{c@chapter}{\newfloat{codelisting}{h}{lop}}{\newfloat{codelisting}{h}{lop}[chapter]}
\floatname{codelisting}{Listing}

\makeatother
\makeatletter
\@ifpackageloaded{caption}{}{\usepackage{caption}}
\@ifpackageloaded{subcaption}{}{\usepackage{subcaption}}
\makeatother
\ifLuaTeX
  \usepackage{selnolig}  % disable illegal ligatures
\fi
\usepackage{bookmark}

\IfFileExists{xurl.sty}{\usepackage{xurl}}{} % add URL line breaks if available
\hypersetup{
  pdftitle={The landscapemetrics and motif packages for measuring landscape patterns and processes},
  pdfauthor={Jakub Nowosad; Maximilian H.K. Hesselbarth},
  colorlinks=true,
  linkcolor={blue},
  filecolor={Maroon},
  citecolor={Blue},
  urlcolor={Blue},
  pdfcreator={LaTeX via pandoc}}

\title{The landscapemetrics and motif packages for measuring landscape
patterns and processes}

  \author{Jakub Nowosad}
            \affil{%
                  Adam Mickiewicz University, Institute of Geoecology
                  and Geoinformation, Poznan, Poland
              }
        \author{Maximilian H.K. Hesselbarth}
            \affil{%
                  International Institute for Applied Systems Analysis,
                  Biodiversity Ecology and Conservation Research Group,
                  Laxenburg, Austria
              }
      
\date{}
\begin{document}
\maketitle

\emph{Abstract:}

This book chapter emphasizes the significance of categorical raster data
in ecological studies, specifically land use or land cover (LULC) data,
and highlights the pivotal role of landscape metrics and pattern-based
spatial analysis in comprehending environmental patterns and their
dynamics. It explores the usage of R packages, particularly
\textbf{landscapemetrics} and \textbf{motif}, for quantifying and
analyzing landscape patterns using LULC data from three distinct
European regions. It showcases the computation, visualization, and
comparison of landscape metrics, while also addressing additional
features such as patch value extraction, sub-region sampling, and moving
window computation. Furthermore, the chapter delves into the intricacies
of pattern-based spatial analysis, explaining how spatial signatures are
computed and how the \textbf{motif} package facilitates comparisons and
clustering of landscape patterns. The chapter concludes by discussing
the potential of customization and expansion of the presented tools.


\subsection{Introduction}\label{introduction}

Categorical raster data, such as ones representing land use or land
cover (LULC), is often used in ecological studies (Fassnacht et al.
2006; Wulder et al. 2018; Chandra Pandey et al. 2021). This data is
typically derived using remote sensing data, for example, satellite
images in combination with statistical learning (Talukdar et al. 2020;
Wang et al. 2022; Wang and Mountrakis 2023). LULC describes the spatial
distribution of different anthropogenic uses of land or the natural land
cover in a landscape. Thus, it provides information about human
activities (usage) and natural features (physical material at the
surface) in a given area within a specific time frame (Fisher et al.
2005). This information can then be used to better understand the
processes and changes taking place in the landscape, such as
urbanization (Fu and Weng 2016), deforestation (Floreano and De Moraes
2021), or the spread of invasive species (Manzoor et al. 2021).

From a computational perspective, LULC data can be conceptualized as a
collection of cells organized in a regular grid, with each cell assigned
to a specific category (With 2019). This framework enables us to analyze
the data in terms of its two fundamental components: the composition,
referring to the number of cells for each category, and the
configuration, pertaining to the spatial arrangement of cells within
each category (Riitters 2019). These components jointly create spatial
patterns of categorical raster data, or landscape patterns, in short,
which are commonly used to characterize the structure of landscapes.
Furthermore, additional characteristics of a categorical raster, for
example, the diversity of categories, connectivity, or patch shape, can
also be considered (Gustafson 1998; Uuemaa et al. 2013).

The aim of this chapter is to provide an overview of the methods and
tools that allow to quantify and analyze landscape patterns. We will
focus on the methods that are mainly implemented in two R packages:
\textbf{landscapemetrics} (Hesselbarth et al. 2019) and \textbf{motif}
(Nowosad 2021). The \textbf{landscapemetrics} package allows to compute
a collection of landscape metrics, which quantify the composition and
configuration of categorical raster data. The package also provides a
set of additional functions, making it possible to visualize the results
of the calculations, extract values of patches, sample metrics within
sub-regions, or apply a moving window computation. Thus, its main
purpose is to provide a set of tools to describe landscape patterns. The
\textbf{motif} package, on the other hand, is focused on the analysis of
these patterns. It allows to compare landscape patterns between
different times, search for areas with resembling landscape patterns, or
cluster areas with similar landscape patterns.

In this chapter, we will consistently utilize a set of example data.
These data consist of spatial raster representing the LULC of three
distinct regions in Europe: the Centre-Val de Loire region in France,
the Noord-Brabant region in the Netherlands, and the Norra Mellansverige
region in Sweden. We obtained the data from the Copernicus mission 2018
(European Environment Agency (EEA) (2023), available at
\textless https://land.copernicus.eu/), and cropped them to the
respective regions. Additionally, all original 45 LULC classes were
re-classified into five general classes (i.e., ``urban'',
``agriculture'', ``vegetation'', ``marshes'', ``water'') to simplify all
examples. To reproduce the results presented in this chapter, you can
download the corresponding data and code from the GitHub repository at
\url{https://github.com/Nowosad/landscapemetrics_motif_2024}.

\subsection{Landscape metrics}\label{landscape-metrics}

Landscape metrics allow to quantify the composition (i.e., amount) of
and configuration (i.e., arrangement) of spatial landscape
characteristics within the context of categorical raster data (Gustafson
1998, 2019; Uuemaa et al. 2013). The advantages of the landscape metrics
approach include their easy application, interpretation, and
communication, especially in landscapes with clear distinctions between
classes (Lausch et al. 2015). Thus, they are a common tool in landscape
and spatial ecology (Kupfer 2012; Lausch et al. 2015; Frazier and Kedron
2017). Even though the first conceptual developments date back to the
1980s (Turner 1989; Gustafson 1998, 2019), the application of landscape
metrics gained popularity with the release of the FRAGSTATS software
(McGarigal et al. 2012) in 1995, which allows to calculate an extensive
collection of metrics (Kupfer 2012; Gustafson 2019).

While FRAGSTATS pioneered the field and made landscape metrics available
to many scientists using a graphical user interface, more recently, the
\textbf{landscapemetrics} R package (Hesselbarth et al. 2019) allows to
calculate a large collection of metrics using the R programming
environment, which is popular among ecologists (Lai et al. 2019;
Hesselbarth et al. 2021). The \textbf{landscapemetrics} package is based
on the \textbf{terra} R package (Hijmans 2021), but additionally
supports raster objects from the \textbf{stars} (Pebesma 2019) and
\textbf{raster} (Hijmans 2019) packages. Furthermore, some functionality
also supports vector data (e.g., sample points) using the \textbf{sf}
package (Pebesma 2018).

\begin{figure}

\centering{

\includegraphics[width=4.8in,height=\textheight]{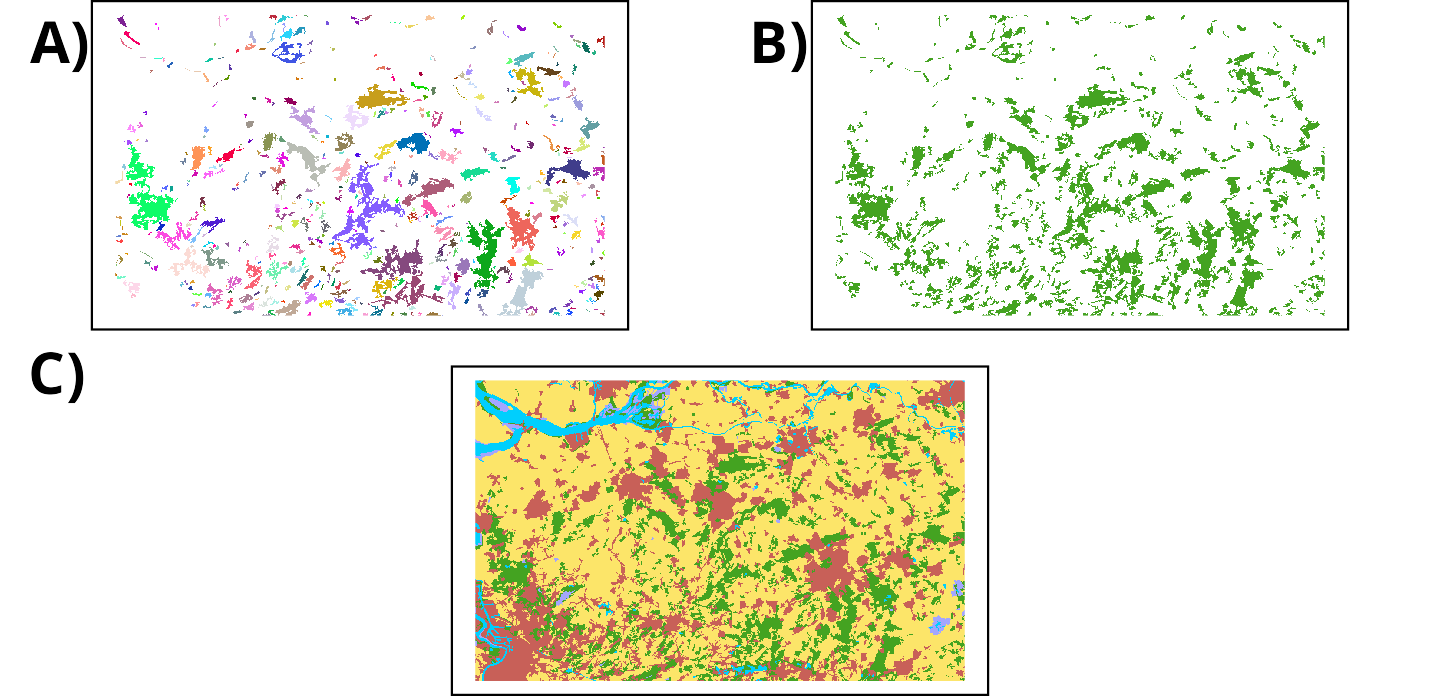}

}

\caption{\label{fig-example-scale}A) Patch level: patches of an
exemplary class indicated by color, i.e., all connected cells belonging
to the same LULC class. B) Class level: exemplary class indicated by
color, i.e., all patches belonging to the same LULC class. C) Landscape
level: exemplary landscape including all LULC classes}

\end{figure}%

Following McGarigal et al.~(2012), landscape metrics can be classified
depending on the characteristics they describe, namely: area and edge,
shape, core area, contrast, aggregation, and diversity metrics.
Furthermore, landscape metrics can be calculated on patch-, class- and
landscape-scale. While patch level metrics describe each patch, i.e.,
contiguous cells belonging to the same class as identified by a
connected components labeling algorithm, class level metrics describe
all patches belonging to the same class, and finally, landscape level
metrics describe the entire landscape Figure~\ref{fig-example-scale}. In
this chapter, we will demonstrate how to calculate all these types of
landscape metrics.

In order to use all functionality of the \textbf{landscapemetrics}
package and to pre- and post-process data, we need to load several
packages related to spatial analysis and data wrangling. This includes
the \textbf{terra} and \textbf{sf} packages for methods related to
predominantly raster and vector data, respectively. Furthermore, the
\textbf{dplyr} (Wickham et al. 2019) and \textbf{tidyr} (Wickham et al.
2023) packages provide a grammar for data manipulation and cleaning,
while \textbf{ggplot2} (Wickham 2016) is a data visualization package.
Additionally, we are creating a color scheme for all maps we are going
to create in later examples.

\begin{Shaded}
\begin{Highlighting}[]
\FunctionTok{library}\NormalTok{(landscapemetrics)}
\FunctionTok{library}\NormalTok{(terra)}
\FunctionTok{library}\NormalTok{(sf)}
\FunctionTok{library}\NormalTok{(dplyr)}
\FunctionTok{library}\NormalTok{(tidyr)}
\FunctionTok{library}\NormalTok{(ggplot2)}
\NormalTok{color\_scale }\OtherTok{\textless{}{-}} \FunctionTok{c}\NormalTok{(}\AttributeTok{urban=}\StringTok{"\#C86058"}\NormalTok{, }\AttributeTok{agriculture=}\StringTok{"\#FCE569"}\NormalTok{, }
                 \AttributeTok{vegetation=}\StringTok{"\#44A321"}\NormalTok{, }\AttributeTok{marshes=}\StringTok{"\#A3A6FF"}\NormalTok{,}
                 \AttributeTok{water=}\StringTok{"\#00CFFD"}\NormalTok{, }\AttributeTok{nodata=}\StringTok{"\#666666"}\NormalTok{)}
\end{Highlighting}
\end{Shaded}

Next, we import the three raster objects using the \textbf{terra}
package representing the LULC data of the three exemplary regions as
\texttt{SpatRaster} objects.

\begin{Shaded}
\begin{Highlighting}[]
\NormalTok{france }\OtherTok{\textless{}{-}} \FunctionTok{rast}\NormalTok{(}\StringTok{"data/raster\_france.tif"}\NormalTok{)}
\NormalTok{netherlands }\OtherTok{\textless{}{-}} \FunctionTok{rast}\NormalTok{(}\StringTok{"data/raster\_netherlands.tif"}\NormalTok{)}
\NormalTok{sweden }\OtherTok{\textless{}{-}} \FunctionTok{rast}\NormalTok{(}\StringTok{"data/raster\_sweden.tif"}\NormalTok{)}
\end{Highlighting}
\end{Shaded}

Before calculating metrics, we need to check if the input raster
fulfills certain requirements of the \textbf{landscapemetrics} package.
For this, we are using the \texttt{check\_landscape()} function. The
function checks if the coordinate reference system (CRS) is projected
(i.e., using Cartesian coordinates on a planar surface), the units of
the map, and the number of classes. If some of the checks fail, we are
still able to calculate all metrics, however, especially metrics relying
on distances might not be correct or hard to interpret, e.g., due to the
decimal degrees units of most geographic CRS (on the contrary metric
distance units of most projected CRS). In this case, the function will
return a corresponding warning.

\begin{Shaded}
\begin{Highlighting}[]
\FunctionTok{check\_landscape}\NormalTok{(france)}
\end{Highlighting}
\end{Shaded}

\begin{verbatim}
  layer       crs units   class n_classes OK
1     1 projected     m integer         5  v
\end{verbatim}

The input raster, \texttt{france}, has a projected CRS with units in
meters and 5 discrete classes in total. Because the raster does not
violate any of the checks, the function returns an OK check mark.

\subsubsection{Deriving landscape
metrics}\label{deriving-landscape-metrics}

All functions to calculate metrics start with the prefix \texttt{lsm\_},
followed by an abbreviation for the level (either \texttt{p} = patch,
\texttt{c} = class, \texttt{l} = landscape), and finally, an
abbreviation of the metric. We can get an overview of all 133 metrics
that are currently provided by the \textbf{landscapemetrics} package
using the \texttt{list\_lsm()} function. Additionally, we can use the
function to display only a certain subset of metrics based on type or
level.

\begin{Shaded}
\begin{Highlighting}[]
\FunctionTok{list\_lsm}\NormalTok{()}
\end{Highlighting}
\end{Shaded}

\begin{verbatim}
# A tibble: 133 x 5
  metric name                          type                 level function_name
  <chr>  <chr>                         <chr>                <chr> <chr>        
1 area   patch area                    area and edge metric patch lsm_p_area   
2 cai    core area index               core area metric     patch lsm_p_cai    
3 circle related circumscribing circle shape metric         patch lsm_p_circle 
# i 130 more rows
\end{verbatim}

\begin{Shaded}
\begin{Highlighting}[]
\FunctionTok{list\_lsm}\NormalTok{(}\AttributeTok{type=}\StringTok{"area and edge metric"}\NormalTok{, }\AttributeTok{level=}\StringTok{"class"}\NormalTok{)}
\end{Highlighting}
\end{Shaded}

\begin{verbatim}
# A tibble: 11 x 5
  metric  name       type                 level function_name
  <chr>   <chr>      <chr>                <chr> <chr>        
1 area_cv patch area area and edge metric class lsm_c_area_cv
2 area_mn patch area area and edge metric class lsm_c_area_mn
3 area_sd patch area area and edge metric class lsm_c_area_sd
# i 8 more rows
\end{verbatim}

To calculate a singular metric on patch-, class- or landscape-level, we
simply use the function name and provide the input raster.

\begin{Shaded}
\begin{Highlighting}[]
\NormalTok{df\_p\_shape }\OtherTok{\textless{}{-}} \FunctionTok{lsm\_p\_shape}\NormalTok{(france)}
\NormalTok{df\_c\_area }\OtherTok{\textless{}{-}} \FunctionTok{lsm\_c\_area\_mn}\NormalTok{(france)}
\NormalTok{df\_l\_lpi }\OtherTok{\textless{}{-}} \FunctionTok{lsm\_l\_lpi}\NormalTok{(france)}
\end{Highlighting}
\end{Shaded}

Thus, to calculate the shape index of each patch, the mean patch area of
each class \emph{i}, or the largest patch index of the landscape, we use
the three functions shown above. The returned data frames always include
the same columns regardless of the specified metric.

\begin{Shaded}
\begin{Highlighting}[]
\NormalTok{df\_l\_lpi}
\end{Highlighting}
\end{Shaded}

\begin{verbatim}
# A tibble: 1 x 6
  layer level     class    id metric value
  <int> <chr>     <int> <int> <chr>  <dbl>
1     1 landscape    NA    NA lpi     65.1
\end{verbatim}

These are an identifier of the raster layer, the level of the calculated
metric, the class identifier, the patch identifier, the name of the
metric, and finally, the metric value. If any of the columns are not
applicable (e.g., the patch identifier on the landscape level), the
column will contain \texttt{NA} values. This allows us to combine
different data frames or apply the same post-processing workflow, even
for different metrics and/or levels. For example, we can use the
\texttt{rbind()} function to combine all three results into a single
data frame and use functions from the \textbf{dplyr} package to
calculate the minimum and maximum value for each metric.

\begin{Shaded}
\begin{Highlighting}[]
\NormalTok{result\_combined }\OtherTok{\textless{}{-}} \FunctionTok{rbind}\NormalTok{(df\_p\_shape, df\_c\_area, df\_l\_lpi)}
\NormalTok{result\_range }\OtherTok{\textless{}{-}} \FunctionTok{group\_by}\NormalTok{(result\_combined, metric) }\SpecialCharTok{|\textgreater{}} 
  \FunctionTok{summarise}\NormalTok{(}\AttributeTok{min=}\FunctionTok{min}\NormalTok{(value), }\AttributeTok{max=}\FunctionTok{max}\NormalTok{(value))}
\NormalTok{result\_range}
\end{Highlighting}
\end{Shaded}

\begin{verbatim}
# A tibble: 3 x 3
  metric    min    max
  <chr>   <dbl>  <dbl>
1 area_mn  64.4 3049. 
2 lpi      65.1   65.1
3 shape     1     84.5
\end{verbatim}

If we want to derive more than a few metrics, we can also use the
\texttt{calculate\_lsm()} function, which allows to calculate several
metrics with only one function call. However, as pointed out by other
authors (Gustafson 2019), ``metric fishing'' is an issue, i.e.,
computing as many metrics as possible and searching for any signal.
Thus, we strongly advise not to calculate all metrics, but rather select
metrics based on research questions and hypotheses. The
\texttt{calculate\_lsm()} function has many options for how to specify
metrics, e.g., a vector of function names or the types and levels of
metrics that can be specified.

\begin{Shaded}
\begin{Highlighting}[]
\NormalTok{df\_area }\OtherTok{\textless{}{-}} \FunctionTok{calculate\_lsm}\NormalTok{(france, }\AttributeTok{what=}\FunctionTok{c}\NormalTok{(}\StringTok{"lsm\_p\_area"}\NormalTok{,}
                                        \StringTok{"lsm\_c\_area\_mn"}\NormalTok{, }\StringTok{"lsm\_l\_ta"}\NormalTok{))}
\NormalTok{df\_aggr\_lsm }\OtherTok{\textless{}{-}} \FunctionTok{calculate\_lsm}\NormalTok{(france, }\AttributeTok{level=}\StringTok{"landscape"}\NormalTok{,}
                                     \AttributeTok{type=}\StringTok{"aggregation metric"}\NormalTok{)}
\end{Highlighting}
\end{Shaded}

For many metrics, we can specify further arguments, such as the
direction of the connected labeling algorithm, which cells are
considered core or edge cells, or the potential number of maximum
classes. To see what arguments are available, we can have a look at the
help page of a function, e.g., \texttt{?lsm\_p\_core}, but all arguments
of a specific metric function can also be used with
\texttt{calculate\_lsm()}. For example, we can change the
\texttt{directions} of the connected labeling algorithm and calculate
the number of patches on the class level. Comparing the results using a
workflow including the \textbf{dplyr} package, we can see that the
number of patches is larger when using ``rook's'' (4-neighborhood)
instead of ``queen's'' rule (8-neighborhood; the default).

\begin{Shaded}
\begin{Highlighting}[]
\NormalTok{df\_np\_queen }\OtherTok{\textless{}{-}} \FunctionTok{lsm\_c\_np}\NormalTok{(france, }\AttributeTok{directions=}\DecValTok{8}\NormalTok{)}
\NormalTok{df\_np\_rook }\OtherTok{\textless{}{-}} \FunctionTok{lsm\_c\_np}\NormalTok{(france, }\AttributeTok{directions=}\DecValTok{4}\NormalTok{)}
\NormalTok{df\_comparison }\OtherTok{\textless{}{-}} \FunctionTok{full\_join}\NormalTok{(}\AttributeTok{x=}\NormalTok{df\_np\_queen, }\AttributeTok{y=}\NormalTok{df\_np\_rook, }
                           \AttributeTok{by=}\FunctionTok{c}\NormalTok{(}\StringTok{"layer"}\NormalTok{, }\StringTok{"level"}\NormalTok{, }\StringTok{"class"}\NormalTok{, }\StringTok{"id"}\NormalTok{, }\StringTok{"metric"}\NormalTok{), }
                           \AttributeTok{suffix=}\FunctionTok{c}\NormalTok{(}\StringTok{".queen"}\NormalTok{, }\StringTok{".rook"}\NormalTok{)) }\SpecialCharTok{|\textgreater{}} 
  \FunctionTok{mutate}\NormalTok{(}\AttributeTok{value.diff=}\FunctionTok{abs}\NormalTok{(value.queen }\SpecialCharTok{{-}}\NormalTok{ value.rook))}
\NormalTok{df\_comparison}
\end{Highlighting}
\end{Shaded}

\begin{verbatim}
# A tibble: 5 x 8
  layer level class    id metric value.queen value.rook value.diff
  <int> <chr> <int> <int> <chr>        <dbl>      <dbl>      <dbl>
1     1 class     1    NA np            3231       4589       1358
2     1 class     2    NA np            1541       3592       2051
3     1 class     3    NA np            5512       9242       3730
# i 2 more rows
\end{verbatim}

\begin{figure}

\centering{

\includegraphics[width=4.8in,height=\textheight]{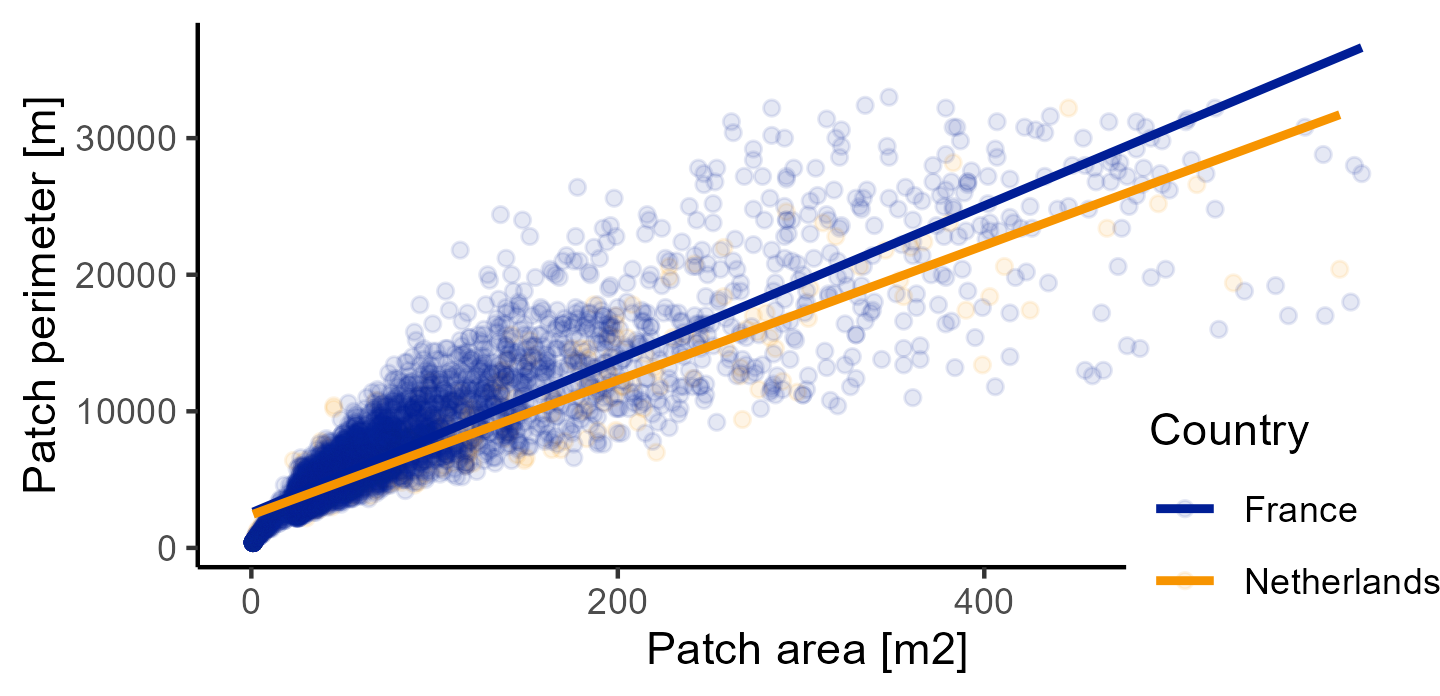}

}

\caption{\label{fig-workflow}Relationship between the patch area and
patch perimeter in the Centre-Val de Loire region in France (blue) and
Noord-Brabant region in the Netherlands (orange)}

\end{figure}

We can also use the \texttt{calculate\_lsm()} function (or any other
\texttt{lsm\_} function) to calculate metrics of several raster layers
or objects simultaneously. For this, we need to use a
\texttt{SpatRaster} with several layers or a list of
\texttt{SpatRaster}'s as input. We are using the later example to
calculate the perimeter and area of each patch for two regions in
Europe. For this example, we only want to keep the vegetation class,
i.e., class 3. Additionally, because some of the patches within each
landscape are disproportional large, we are going to use the 95\%
quantiles of each region and metric to remove some larger values. Next,
we are re-labeling the unique layer identification column using the
country names. Last, we are reshaping the data frame from a long to a
wider format to produce a \textbf{ggplot2} figure in the final step
Figure~\ref{fig-workflow}.

\begin{Shaded}
\begin{Highlighting}[]
\NormalTok{df\_perim\_core }\OtherTok{\textless{}{-}} \FunctionTok{list}\NormalTok{(}\AttributeTok{nl=}\NormalTok{netherlands, }\AttributeTok{fr=}\NormalTok{france) }\SpecialCharTok{|\textgreater{}} 
  \FunctionTok{calculate\_lsm}\NormalTok{(}\AttributeTok{what=}\FunctionTok{c}\NormalTok{(}\StringTok{"lsm\_p\_perim"}\NormalTok{, }\StringTok{"lsm\_p\_area"}\NormalTok{)) }\SpecialCharTok{|\textgreater{}} 
  \FunctionTok{filter}\NormalTok{(class }\SpecialCharTok{\%in\%} \DecValTok{3}\NormalTok{) }\SpecialCharTok{|\textgreater{}} 
  \FunctionTok{mutate}\NormalTok{(}\AttributeTok{layer=}\FunctionTok{case\_when}\NormalTok{(layer}\SpecialCharTok{==}\DecValTok{1}\SpecialCharTok{\textasciitilde{}}\StringTok{"Netherlands"}\NormalTok{, }
\NormalTok{                         layer}\SpecialCharTok{==}\DecValTok{2}\SpecialCharTok{\textasciitilde{}}\StringTok{"France"}\NormalTok{)) }\SpecialCharTok{|\textgreater{}} 
  \FunctionTok{pivot\_wider}\NormalTok{(}\AttributeTok{names\_from=}\NormalTok{metric, }\AttributeTok{values\_from=}\NormalTok{value) }\SpecialCharTok{|\textgreater{}} 
  \FunctionTok{filter}\NormalTok{(area}\SpecialCharTok{\textless{}=}\FunctionTok{quantile}\NormalTok{(area, }\AttributeTok{probs=}\FloatTok{0.95}\NormalTok{) }\SpecialCharTok{\&}
\NormalTok{         perim}\SpecialCharTok{\textless{}=}\FunctionTok{quantile}\NormalTok{(perim, }\AttributeTok{probs=}\FloatTok{0.95}\NormalTok{))}
\FunctionTok{ggplot}\NormalTok{(}\AttributeTok{data=}\NormalTok{df\_perim\_core, }\FunctionTok{aes}\NormalTok{(}\AttributeTok{x=}\NormalTok{area, }\AttributeTok{y=}\NormalTok{perim, }\AttributeTok{color=}\NormalTok{layer)) }\SpecialCharTok{+}
  \FunctionTok{geom\_point}\NormalTok{(}\AttributeTok{alpha=}\FloatTok{0.1}\NormalTok{) }\SpecialCharTok{+}
  \FunctionTok{geom\_smooth}\NormalTok{(}\AttributeTok{se=}\ConstantTok{FALSE}\NormalTok{, }\AttributeTok{method=}\StringTok{"lm"}\NormalTok{, }\AttributeTok{formula=}\StringTok{"y \textasciitilde{} x"}\NormalTok{) }\SpecialCharTok{+}
  \FunctionTok{scale\_color\_manual}\NormalTok{(}\AttributeTok{name=}\StringTok{"Country"}\NormalTok{, }
                     \AttributeTok{values=}\FunctionTok{c}\NormalTok{(}\AttributeTok{Netherlands=}\StringTok{"\#F79400"}\NormalTok{,}
                              \AttributeTok{France=}\StringTok{"\#001E96"}\NormalTok{)) }\SpecialCharTok{+} 
  \FunctionTok{labs}\NormalTok{(}\AttributeTok{x=}\StringTok{"Patch area [m2]"}\NormalTok{, }\AttributeTok{y=}\StringTok{"Patch perimeter [m]"}\NormalTok{) }\SpecialCharTok{+} 
  \FunctionTok{theme\_classic}\NormalTok{() }\SpecialCharTok{+} \FunctionTok{theme}\NormalTok{(}\AttributeTok{legend.position =} \FunctionTok{c}\NormalTok{(}\FloatTok{0.9}\NormalTok{, }\FloatTok{0.1}\NormalTok{))}
\end{Highlighting}
\end{Shaded}

This example also demonstrates how easily the \textbf{landscapemetrics}
package can be incorporated into larger workflows. As expected, there is
a strong relationship between the patch area and perimeter. Logically,
as the patch area also increases the perimeter increases and this trend
is comparable between the two regions.

\subsubsection{Visualization of landscape
metrics}\label{visualization-of-landscape-metrics}

\begin{figure}

\centering{

\includegraphics[width=4.8in,height=\textheight]{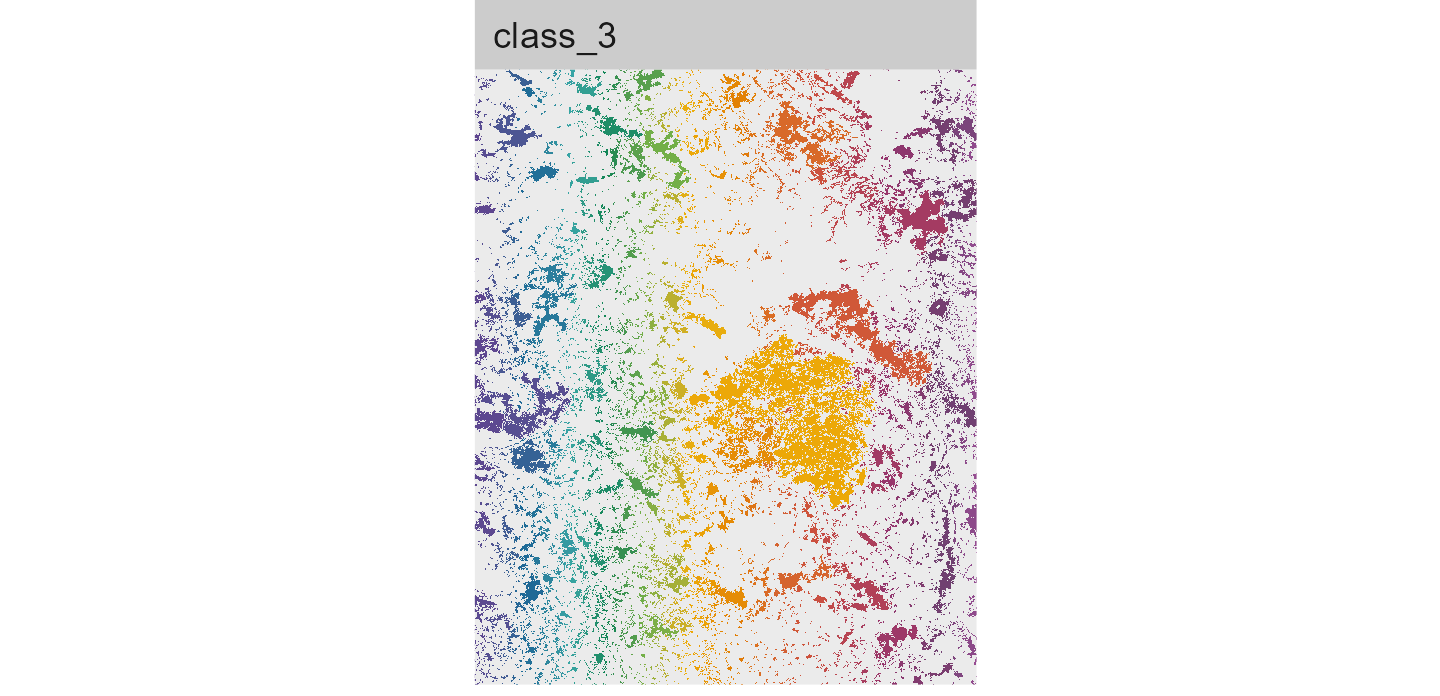}

}

\caption{\label{fig-show-patches}Patches of the natural vegetation class
in the Centre-Val de Loire region in France}

\end{figure}

We can also use built-in functions to visualize the landscapes or
calculated metrics. All visualization functions start with the
\texttt{show\_} prefix and return a list storing \textbf{ggplot2}
objects. This is done to be type-stable (i.e., always returning the same
data type) regardless of the number of raster objects/layers or selected
metrics. First, we can visualize all patches, i.e., all connected cells
as specified by the \texttt{directions} argument of the connected
components labeling algorithm. We can show the patches of the entire
landscape (using the argument \texttt{class="global"}), or only patches
of specific classes (e.g., using \texttt{class=3}). Here, we show all
patches of the vegetation class in the French region
Figure~\ref{fig-show-patches}.

\begin{Shaded}
\begin{Highlighting}[]
\FunctionTok{show\_patches}\NormalTok{(france, }\AttributeTok{class=}\DecValTok{3}\NormalTok{)}
\end{Highlighting}
\end{Shaded}

\begin{figure}

\centering{

\includegraphics[width=4.8in,height=\textheight]{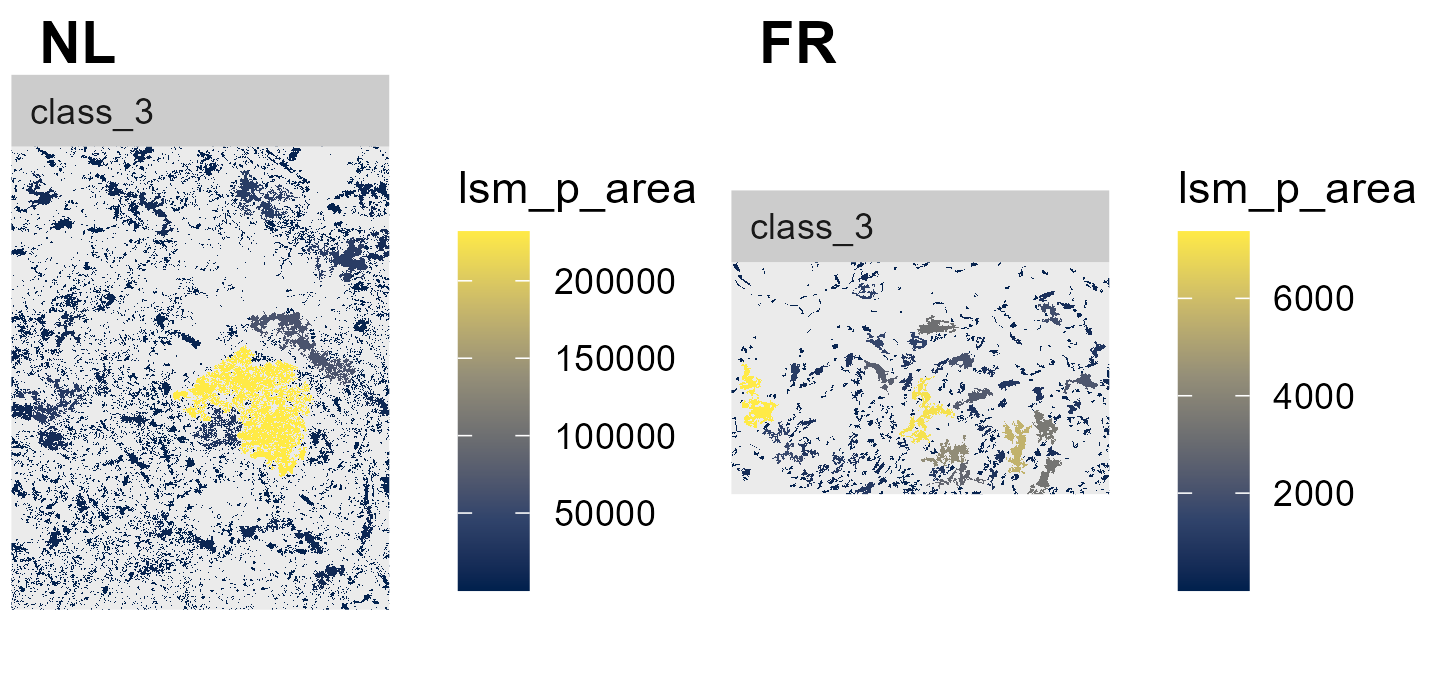}

}

\caption{\label{fig-show-lsm}Visualization of the patch area for the
natural vegetation LULC class in the Centre-Val de Loire region in
France (left) and the Noord-Brabant region in the Netherlands (right)}

\end{figure}

Furthermore, showcasing a more advanced approach, we have the capability
to fill each cell of a patch accordingly to a specific landscape metric.
For instance, in the following example, the fill color of each patch in
the vegetation class corresponds to its area. Since we utilize a list to
store two landscapes, the resulting list contains two elements, each
representing a landscape and storing the associated \texttt{ggplot2}
objects. It becomes apparent that the landscape in France is dominated
by a single large patch situated at the center of the region
Figure~\ref{fig-show-lsm}.

\begin{Shaded}
\begin{Highlighting}[]
\NormalTok{list\_gg\_area }\OtherTok{\textless{}{-}} \FunctionTok{list}\NormalTok{(}\AttributeTok{fr=}\NormalTok{france, }\AttributeTok{nl=}\NormalTok{netherlands) }\SpecialCharTok{|\textgreater{}} 
  \FunctionTok{show\_lsm}\NormalTok{(}\AttributeTok{class=}\DecValTok{3}\NormalTok{, }\AttributeTok{what=}\StringTok{"lsm\_p\_area"}\NormalTok{)}
\FunctionTok{plot\_grid}\NormalTok{(}\AttributeTok{plotlist=}\NormalTok{list\_gg\_area, }\AttributeTok{ncol=}\DecValTok{2}\NormalTok{, }\AttributeTok{labels=}\FunctionTok{c}\NormalTok{(}\StringTok{"NL"}\NormalTok{, }\StringTok{"FR"}\NormalTok{))}
\end{Highlighting}
\end{Shaded}

\begin{figure}

\centering{

\includegraphics[width=4.8in,height=\textheight]{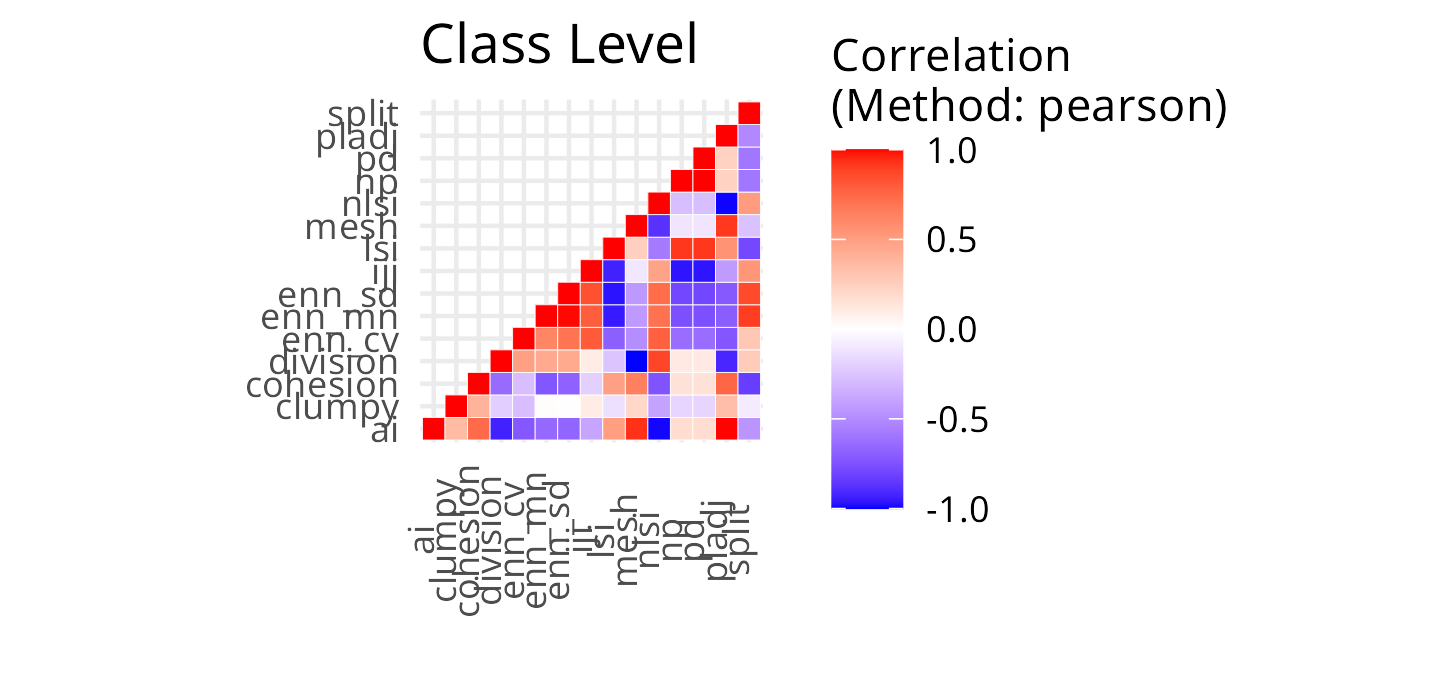}

}

\caption{\label{fig-show-corr}Correlation matrix of all class level
metrics for the Noord-Brabant region in the Netherlands}

\end{figure}

Correlation between several metrics can be an issue (Cushman et al.
2008; Schindler et al. 2008; Nowosad and Stepinski 2018), especially in
combination with the previously discussed practice of ``metric
fishing''. Thus, we can use the \texttt{show\_correlation()} function to
check correlations between metrics that were previously calculated using
either \texttt{calculate\_lsm()} or singular metric functions
Figure~\ref{fig-show-corr}.

\begin{Shaded}
\begin{Highlighting}[]
\NormalTok{class\_metrics }\OtherTok{\textless{}{-}} \FunctionTok{calculate\_lsm}\NormalTok{(netherlands, }\AttributeTok{level=}\StringTok{"class"}\NormalTok{,}
                               \AttributeTok{type=}\StringTok{"aggregation metric"}\NormalTok{)}
\FunctionTok{show\_correlation}\NormalTok{(class\_metrics)}
\end{Highlighting}
\end{Shaded}

\subsubsection{Additional features}\label{additional-features}

The \textbf{landscapemetrics} also provides some more advanced
functionality. This includes extracting values of patches, sampling of
metrics within sub-regions, or applying a moving window approach. In
terms of metrics selection or arguments passed on to the metric
functions, all these functions behave similarly to the
\texttt{calculate\_lsm()} function.

\begin{figure}

\centering{

\includegraphics[width=4.8in,height=\textheight]{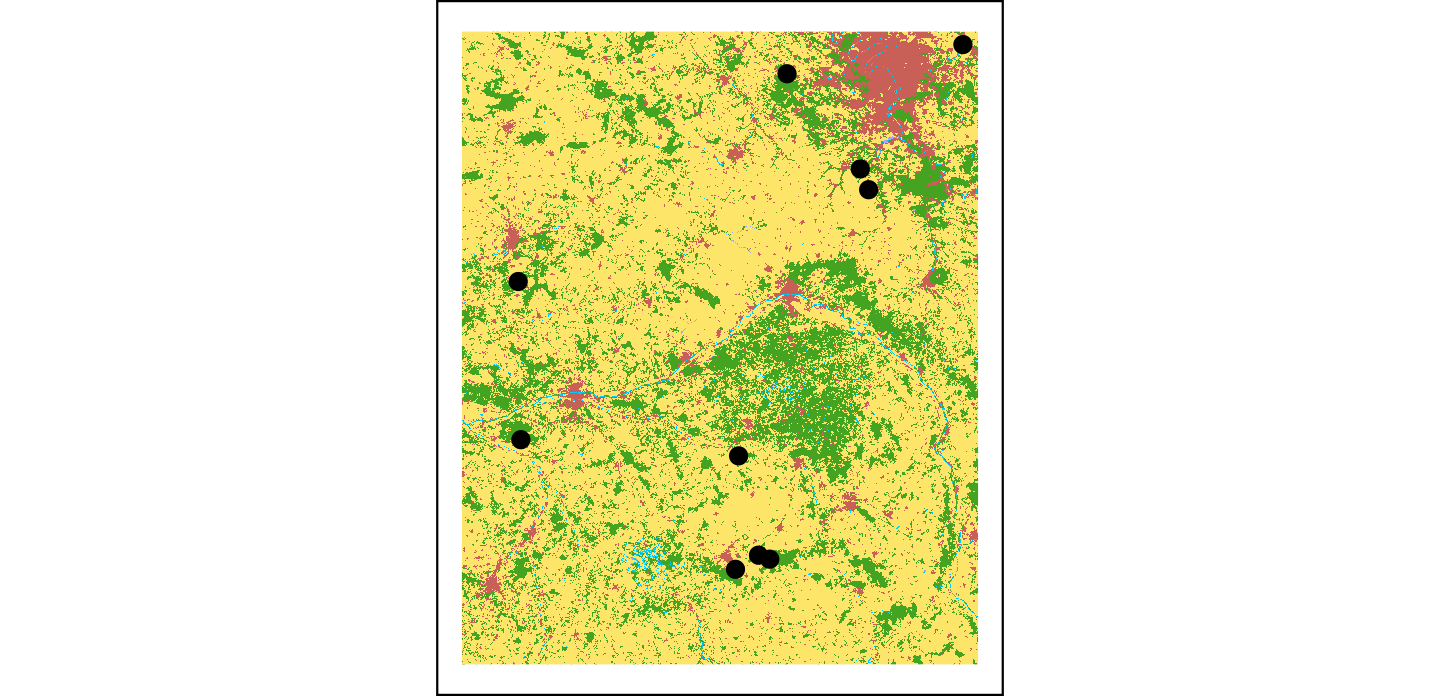}

}

\caption{\label{fig-plot-pts}Location of exemplary sample points used to
extract and sample landscape metrics in the Centre-Val de Loire region
in France}

\end{figure}

In order to demonstrate the extraction and sampling of metrics, we first
need to create sample points. For this, we are going to randomly create
10 points within a region Figure~\ref{fig-plot-pts}.

\begin{Shaded}
\begin{Highlighting}[]
\NormalTok{samplepoints }\OtherTok{\textless{}{-}} \FunctionTok{spatSample}\NormalTok{(france, }\AttributeTok{size=}\DecValTok{10}\NormalTok{, }\AttributeTok{as.points=}\ConstantTok{TRUE}\NormalTok{) }\SpecialCharTok{|\textgreater{}} 
  \FunctionTok{st\_as\_sf}\NormalTok{()}
\end{Highlighting}
\end{Shaded}

Now, we can extract patch level metrics of all patches that contain a
sample point. For this, we use the \texttt{extract\_lsm()} function. The
resulting data frame has 20 rows, one for each metric and sample point.
It is worth noting that compared to previous result data frames, there
is an additional column named \texttt{extract\_id}, which facilitates
matching the metric results with the respective sample points.

\begin{Shaded}
\begin{Highlighting}[]
\NormalTok{df\_extract }\OtherTok{\textless{}{-}} \FunctionTok{extract\_lsm}\NormalTok{(france, samplepoints,}
                          \AttributeTok{what=}\FunctionTok{c}\NormalTok{(}\StringTok{"lsm\_p\_area"}\NormalTok{, }\StringTok{"lsm\_p\_perim"}\NormalTok{))}
\NormalTok{df\_extract}
\end{Highlighting}
\end{Shaded}

\begin{verbatim}
# A tibble: 20 x 7
  layer level class    id metric  value extract_id
  <int> <chr> <int> <int> <chr>   <dbl>      <int>
1     1 patch     3  5207 area    15202          1
2     1 patch     3  5207 perim  297200          1
3     1 patch     3  7899 area     7652          2
# i 17 more rows
\end{verbatim}

Similarly, we can also sample metrics within a buffer around each sample
point. For this, we need to additionally provide a size of the sampling
buffer and its shape (either a square or circle), but it is also
possible to directly use polygon objects as a sampling buffer.

\begin{Shaded}
\begin{Highlighting}[]
\NormalTok{df\_sample }\OtherTok{\textless{}{-}} \FunctionTok{sample\_lsm}\NormalTok{(france, samplepoints,}
                        \AttributeTok{what=}\FunctionTok{c}\NormalTok{(}\StringTok{"lsm\_c\_pland"}\NormalTok{, }\StringTok{"lsm\_l\_ta"}\NormalTok{), }
                        \AttributeTok{size=}\DecValTok{10000}\NormalTok{, }\AttributeTok{shape=}\StringTok{"circle"}\NormalTok{)}
\FunctionTok{filter}\NormalTok{(df\_sample, percentage\_inside}\SpecialCharTok{\textgreater{}=}\DecValTok{75}\NormalTok{)}
\end{Highlighting}
\end{Shaded}

\begin{verbatim}
# A tibble: 47 x 8
  layer level class    id metric value plot_id percentage_inside
  <int> <chr> <int> <int> <chr>  <dbl>   <int>             <dbl>
1     1 class     1    NA pland   3.54       1              101.
2     1 class     2    NA pland  48.6        1              101.
3     1 class     3    NA pland  46.8        1              101.
# i 44 more rows
\end{verbatim}

The resulting data frame has two additional columns. First, similar to
the metrics extraction, an additional column identifies the sampling
buffer. Second, an additional column stores information about the actual
clipped sampling buffer area. Theoretically, this value should be equal
to 100\%. However, it includes all of the raster cells whose centroids
are inside the selector polygon (Lovelace et al. 2019), and thus the
actual sampling buffer area might be slightly larger or smaller than the
specified area resulting in a deviation from the theoretical value.
Furthermore, sample points at the edge of the landscape might include a
smaller area than specified by the size argument. In these cases, a
warning is returned and these sample plots can potentially be excluded
from further analyses.

Similar to the previously used \texttt{show\_lsm()} function, patch
level metrics can be returned as a raster object in which each cell
stores the metric value of the patch it belongs to. Thus, each cell in
the raster object stores its corresponding metrics value. For this, the
function \texttt{spatialize\_lsm()} can be used. However, in order to be
type-stable, this functions always returns a nested list for each layer
and metric (even if only one layer and/or metric is present). The first
level of the list corresponds to the number of layers, while the second
level of the list corresponds to the number of selected metrics. So we
need to use some indexing in order to get only one raster, e.g., the
fractal dimension index raster, which we now can use for further
analysis, such as calculating kernel density estimates (results not
shown).

\begin{Shaded}
\begin{Highlighting}[]
\NormalTok{list\_shape }\OtherTok{\textless{}{-}} \FunctionTok{spatialize\_lsm}\NormalTok{(netherlands,}
                             \AttributeTok{what=}\FunctionTok{c}\NormalTok{(}\StringTok{"lsm\_p\_shape"}\NormalTok{, }\StringTok{"lsm\_p\_frac"}\NormalTok{))}
\FunctionTok{class}\NormalTok{(list\_shape)}
\FunctionTok{str}\NormalTok{(list\_shape)}
\FunctionTok{values}\NormalTok{(list\_shape}\SpecialCharTok{$}\NormalTok{layer\_1}\SpecialCharTok{$}\NormalTok{lsm\_p\_frac, }\AttributeTok{mat=}\ConstantTok{FALSE}\NormalTok{) }\SpecialCharTok{|\textgreater{}} \FunctionTok{density}\NormalTok{()}
\end{Highlighting}
\end{Shaded}

Last, \textbf{landscapemetrics} provides a moving window approach, i.e.,
metrics can be calculated for the local neighborhood of each focal cell.
For this, the \texttt{window\_lsm()} function can be used. Similar to
the \texttt{focal()} function of the \textbf{terra} package, a local
neighborhood must be specific using a matrix object. As with the
previously discussed functions, metrics can be selected using the type
of metric, or specific metrics using the \texttt{what} argument.
However, currently only landscape level metrics are implemented.
Depending on the size of the landscape and the local neighborhood
matrix, this can be computationally demanding with a long run time.

\begin{Shaded}
\begin{Highlighting}[]
\NormalTok{mat\_window }\OtherTok{\textless{}{-}} \FunctionTok{matrix}\NormalTok{(}\DecValTok{1}\NormalTok{, }\AttributeTok{nrow=}\DecValTok{501}\NormalTok{, }\AttributeTok{ncol=}\DecValTok{501}\NormalTok{)}
\FunctionTok{window\_lsm}\NormalTok{(netherlands, }\AttributeTok{window=}\NormalTok{mat\_window, }\AttributeTok{what=}\StringTok{"lsm\_l\_pr"}\NormalTok{)}
\end{Highlighting}
\end{Shaded}

\subsubsection{Utility functions}\label{utility-functions}

Many of the functions internally used by the \textbf{landscapemetrics}
package might also be useful for users during pre- or post-processing of
raster data or to develop new methods to quantify landscape
characteristics. All of these functions start with the \texttt{get\_}
suffix and usually return a (nested) list to be type-stable for
different inputs.

One of the most fundamental functions of the package is the
\texttt{get\_patches()} function, which returns the patches of each
class using the connected components labeling algorithm. Yet, the result
is a nested list in which the first level refers to the number of raster
layers/objects and the second level stores all patches separated by
class \emph{i}. All other cells not belonging to the current class
\emph{i} are set to \texttt{NA}. In the following example, we are first
getting all patches of the region in French separated by class. Next, we
are using all wetland and water-related classes and combining them into
one \texttt{SpatRaster} object as layers. Last, we use raster algebra to
sum all cells which returns a singular layer in which all cells are
labelled by their cell identifier and all other cells are \texttt{NA}.
This results in a single raster layer combining all patches of one of
the two water-related classes.

\begin{Shaded}
\begin{Highlighting}[]
\NormalTok{list\_patches }\OtherTok{\textless{}{-}} \FunctionTok{get\_patches}\NormalTok{(france)}
\NormalTok{ras\_water }\OtherTok{\textless{}{-}}\NormalTok{ list\_patches}\SpecialCharTok{$}\NormalTok{layer\_1[}\DecValTok{4}\SpecialCharTok{:}\DecValTok{5}\NormalTok{] }\SpecialCharTok{|\textgreater{}} 
  \FunctionTok{rast}\NormalTok{() }\SpecialCharTok{|\textgreater{}} 
  \FunctionTok{sum}\NormalTok{(}\AttributeTok{na.rm=}\ConstantTok{TRUE}\NormalTok{)}
\NormalTok{ras\_water}
\end{Highlighting}
\end{Shaded}

\begin{verbatim}
class       : SpatRaster 
dimensions  : 2860, 2333, 1  (nrow, ncol, nlyr)
resolution  : 100, 100  (x, y)
extent      : 3568800, 3802100, 2619700, 2905700  (xmin, xmax, ymin, ymax)
coord. ref. : ETRS89-extended / LAEA Europe (EPSG:3035) 
source(s)   : memory
name        :   sum 
min value   : 10285 
max value   : 10844 
\end{verbatim}

The \texttt{get\_boundaries()} function returns a raster in which all
boundary/edge cells of a patch are labeled 1 and all core cells are
labeled 0. We define an edge cell as a cell that shares a neighboring
cell with a different LULC value than itself. Usually, we apply the
function to previously connected components labeled landscapes and
receive a list with one element for each layer/object. This list can be
used for further analysis, such as counting the number of edge and core
cells.

\begin{Shaded}
\begin{Highlighting}[]
\NormalTok{ras\_boundaries\_urban }\OtherTok{\textless{}{-}} \FunctionTok{get\_boundaries}\NormalTok{(list\_patches}\SpecialCharTok{$}\NormalTok{layer\_1}\SpecialCharTok{$}\NormalTok{class\_1)}
\FunctionTok{lapply}\NormalTok{(ras\_boundaries\_urban, freq, }\AttributeTok{wide=}\ConstantTok{TRUE}\NormalTok{)}
\end{Highlighting}
\end{Shaded}

\begin{verbatim}
[[1]]
  layer      0      1
1     1 304595 164282
\end{verbatim}

We can use the two functions \texttt{get\_unique\_values()} and
\texttt{get\_adjacencies()} to receive a vector with all unique class
values and their adjacency matrix, respectively. Because both return a
list in order to be type-stable, we can directly index the first element
if we only provide one raster layer/object. Of course, the dimensions of
the adjacency matrix must be identical to the length of the unique class
values given.

\begin{Shaded}
\begin{Highlighting}[]
\NormalTok{vec\_unique }\OtherTok{\textless{}{-}} \FunctionTok{get\_unique\_values}\NormalTok{(sweden)[[}\DecValTok{1}\NormalTok{]]}
\NormalTok{mat\_adjacencies }\OtherTok{\textless{}{-}} \FunctionTok{get\_adjacencies}\NormalTok{(sweden)[[}\DecValTok{1}\NormalTok{]]}
\FunctionTok{dim}\NormalTok{(mat\_adjacencies)}
\end{Highlighting}
\end{Shaded}

\begin{verbatim}
[1] 5 5
\end{verbatim}

\begin{Shaded}
\begin{Highlighting}[]
\FunctionTok{length}\NormalTok{(vec\_unique)}
\end{Highlighting}
\end{Shaded}

\begin{verbatim}
[1] 5
\end{verbatim}

Related, we can use the \texttt{get\_nearestneighbour()} function on a
previously connected component labeled landscape, which returns a data
frame with the minimum Euclidean distances to the nearest patch of the
same class. In the following example, we calculate the minimum distances
between all patches of class 1, i.e., urban areas. There are 1,175
patches in total and the distance, as well as the identifier of the
nearest neighbor, is included. We see that the distance between patches
626 and 650 of the urban class is the largest in the LULC data of
Sweden.

\begin{Shaded}
\begin{Highlighting}[]
\NormalTok{sweden\_cl1 }\OtherTok{\textless{}{-}} \FunctionTok{get\_patches}\NormalTok{(sweden, }\AttributeTok{class=}\DecValTok{1}\NormalTok{)}
\NormalTok{df\_neighbour }\OtherTok{\textless{}{-}}\NormalTok{ sweden\_cl1}\SpecialCharTok{$}\NormalTok{layer\_1}\SpecialCharTok{$}\NormalTok{class\_1 }\SpecialCharTok{|\textgreater{}}
  \FunctionTok{get\_nearestneighbour}\NormalTok{(}\AttributeTok{return\_id=}\ConstantTok{TRUE}\NormalTok{) }\SpecialCharTok{|\textgreater{}} 
  \FunctionTok{arrange}\NormalTok{(}\SpecialCharTok{{-}}\NormalTok{dist)}
\end{Highlighting}
\end{Shaded}

\begin{figure}

\centering{

\includegraphics[width=4.8in,height=\textheight]{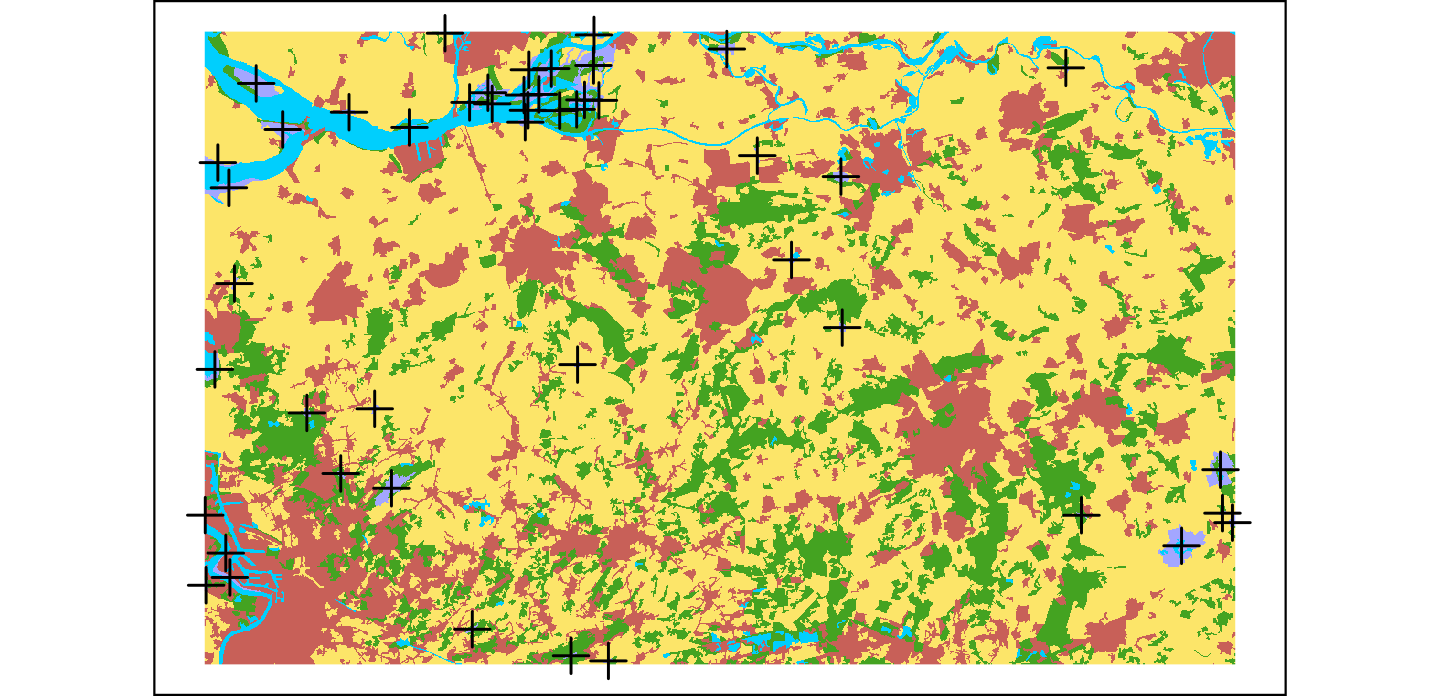}

}

\caption{\label{fig-shape-util}Patch centroids of all marsh patches in
the Noord-Brabant region in the Netherlands}

\end{figure}

Last, there are two utility functions related to the shape of the
patches. First, the function \texttt{get\_circumscribingcircle()}
returns a data frame with the diameter of the smallest circumscribing
circle around each patch and the x- and y-coordinate of it. Similar to
this, the \texttt{get\_centroids()} function returns a data frame with
the coordinates of the centroids of each patch. For this example, we
will filter the data frame to include only the centroids of the marshes
class and create a map plotting the patches as well as all the centroids
Figure~\ref{fig-shape-util}.

\begin{Shaded}
\begin{Highlighting}[]
\NormalTok{df\_circle }\OtherTok{\textless{}{-}} \FunctionTok{get\_circumscribingcircle}\NormalTok{(netherlands)}
\NormalTok{df\_centroids }\OtherTok{\textless{}{-}} \FunctionTok{get\_centroids}\NormalTok{(netherlands) }\SpecialCharTok{|\textgreater{}} 
\NormalTok{  dplyr}\SpecialCharTok{::}\FunctionTok{filter}\NormalTok{(class}\SpecialCharTok{==}\DecValTok{4}\NormalTok{)}
\FunctionTok{as.data.frame}\NormalTok{(netherlands, }\AttributeTok{xy=}\ConstantTok{TRUE}\NormalTok{) }\SpecialCharTok{|\textgreater{}} 
  \FunctionTok{ggplot}\NormalTok{(}\FunctionTok{aes}\NormalTok{(}\AttributeTok{x=}\NormalTok{x, }\AttributeTok{y=}\NormalTok{y)) }\SpecialCharTok{+} 
  \FunctionTok{geom\_raster}\NormalTok{(}\FunctionTok{aes}\NormalTok{(}\AttributeTok{fill=}\FunctionTok{as.factor}\NormalTok{(cover))) }\SpecialCharTok{+}
  \FunctionTok{geom\_point}\NormalTok{(}\AttributeTok{data=}\NormalTok{df\_centroids, }\FunctionTok{aes}\NormalTok{(}\AttributeTok{x=}\NormalTok{x, }\AttributeTok{y=}\NormalTok{y), }\AttributeTok{shape=}\DecValTok{3}\NormalTok{, }\AttributeTok{size=}\FloatTok{2.5}\NormalTok{) }\SpecialCharTok{+}
  \FunctionTok{scale\_fill\_manual}\NormalTok{(}\AttributeTok{values=}\NormalTok{color\_scale) }\SpecialCharTok{+} 
  \FunctionTok{coord\_equal}\NormalTok{() }\SpecialCharTok{+} \FunctionTok{theme\_void}\NormalTok{() }\SpecialCharTok{+} \FunctionTok{theme}\NormalTok{(}\AttributeTok{legend.position=}\StringTok{"none"}\NormalTok{)}
\end{Highlighting}
\end{Shaded}

\subsection{Pattern-based spatial
analysis}\label{pattern-based-spatial-analysis}

Pattern-based spatial analysis is a set of methods allowing for
performing various tasks on landscape patterns, such as comparing
landscape patterns or searching for areas with similar landscape
patterns to the query one (Wickham and Norton 1994; Jasiewicz et al.
2015; Netzel et al. 2018; Nowosad 2021). Instead of treating each cell
independently, this approach focuses on considering local composition
and configuration. Firstly, each area, a group of adjacent cells
encompassing a landscape pattern, is described numerically as spatial
signatures. Next, the similarity between spatial signatures for two
areas can be measured using various dissimilarity measures. Low
dissimilarity values suggest that the two areas have similar composition
and configuration. These ideas allow, for example, to compare spatial
pattern change between an area in time, search for similarities between
areas of interest, or group areas with similar patterns together.
Spatial signatures may also be used directly as new variables, for
example, in machine learning models.

The pattern-based spatial analysis methods are available in the R
package \textbf{motif} (Nowosad 2021), which allows to describe spatial
patterns of one or more categorical raster data for any defined regular
and irregular regions. It accepts spatial raster objects from the
\textbf{terra} (Hijmans 2021) and \textbf{stars} (Pebesma 2019) R
packages as inputs. All \textbf{motif} functions use the \texttt{lsp\_}
(\emph{local spatial pattern}) prefix consistently, allowing users to
find the desired tool quickly. Most functions in the package are based
on a computationally fast C++ code and the software is designed to work
on larger-than-RAM raster datasets.

We need to start by attaching the \textbf{motif} package for
pattern-based spatial analysis, and two main packages for handling
spatial data, \textbf{terra} and \textbf{sf} (Pebesma 2018).

\begin{Shaded}
\begin{Highlighting}[]
\FunctionTok{library}\NormalTok{(motif)}
\FunctionTok{library}\NormalTok{(terra)}
\FunctionTok{library}\NormalTok{(sf)}
\end{Highlighting}
\end{Shaded}

Our main dataset for the following examples will be the spatial raster
representing land cover classes in the Centre-Val de Loire region in
France for the year 2018.

\begin{Shaded}
\begin{Highlighting}[]
\NormalTok{france }\OtherTok{\textless{}{-}} \FunctionTok{rast}\NormalTok{(}\StringTok{"data/raster\_france.tif"}\NormalTok{)}
\FunctionTok{plot}\NormalTok{(france) }\CommentTok{\# result not shown}
\end{Highlighting}
\end{Shaded}

\subsubsection{Spatial signatures}\label{spatial-signatures}

The backbone of pattern-based spatial analysis is the concept of spatial
signatures. A spatial signature is a short numerical descriptor of
landscape patterns featured in a particular area, and represented by a
categorical raster. Landscape patterns can be described in various ways,
and thus many possible spatial signatures exist. That being said, they
usually express the composition and/or configuration (arrangement) of
categorical raster cells (Table~\ref{tbl-spatial-signatures}). It is
also worth noting that spatial signatures can be calculated for a large
area or many smaller subareas divided by regular or irregular zones
(\emph{windows}).

Spatial signatures are derived using the \texttt{lsp\_signature()}
function. We need to provide two arguments to calculate a single spatial
signature: our input categorical raster data (\texttt{x}) and the
spatial signature type (\texttt{type}). The most basic spatial signature
is \texttt{"composition"}: it is a share of cells of each category in a
given area.

\begin{Shaded}
\begin{Highlighting}[]
\NormalTok{france\_comp }\OtherTok{\textless{}{-}} \FunctionTok{lsp\_signature}\NormalTok{(france, }\AttributeTok{type=}\StringTok{"composition"}\NormalTok{)}
\NormalTok{france\_comp}
\end{Highlighting}
\end{Shaded}

\begin{verbatim}
# A tibble: 1 x 3
     id na_prop signature    
* <int>   <dbl> <list>       
1     1       0 <dbl [1 x 5]>
\end{verbatim}

The \texttt{lsp\_signature()} function always returns a data frame with
three columns: a unique identifier (\texttt{id}), the proportion of
cells in an area with missing values (\texttt{na\_prop}), and the
calculated signature (\texttt{signature}). This last column is a list
containing a single signature per area of interest. We can see the
obtained spatial signature by accessing the first element of the
\texttt{signature} list.

\begin{Shaded}
\begin{Highlighting}[]
\NormalTok{france\_comp}\SpecialCharTok{$}\NormalTok{signature[[}\DecValTok{1}\NormalTok{]]}
\end{Highlighting}
\end{Shaded}

\begin{verbatim}
              1         2         3            4         5
[1,] 0.07027133 0.7041287 0.2184272 0.0004055524 0.0067673
\end{verbatim}

It shows that \textasciitilde70\% of the area is covered by agriculture
(\texttt{2}), \textasciitilde22\% by vegetation (\texttt{3}), and
\textasciitilde7\% by the urban class (\texttt{1}).

While the \texttt{"composition"} signature is straightforward to
understand, it does not encompass the spatial arrangement of the
categories: it does not tell if a given category represents one large
continuous area or if it is distributed as many small patches. Thus,
more complex and informative signatures should be used in most cases.
For instance, when dealing with a single categorical raster, a
co-occurrence vector (``cove'') serves as an appropriate choice for such
a signature.

The co-occurrence vector is a spatial signature that encapsulates both
the composition and configuration of raster classes in a given area
(Haralick et al. 1973; Jasiewicz et al. 2015; Nowosad and Stepinski
2021). This is achieved by counting not how many times each class occurs
but how often a pixel with a given class is adjacent to another pixel of
some class. Thus, the result is a long vector (15 elements in this
example) that counts how many times pairs of pixels with given classes
are adjacent, e.g., urban cells are adjacent to other urban cells, and
how many times urban cells are adjacent to, for example, vegetation
cells. In general, you can think of spatial signatures as a way to
compress information: our original raster has 6,672,380 values, which we
now compressed into a vector of only 15 elements.

\begin{Shaded}
\begin{Highlighting}[]
\NormalTok{france\_cove }\OtherTok{\textless{}{-}} \FunctionTok{lsp\_signature}\NormalTok{(france, }\AttributeTok{type=}\StringTok{"cove"}\NormalTok{)}
\NormalTok{france\_cove}
\end{Highlighting}
\end{Shaded}

\begin{verbatim}
# A tibble: 1 x 3
     id na_prop signature     
* <int>   <dbl> <list>        
1     1       0 <dbl [1 x 15]>
\end{verbatim}

As shown above, spatial signatures may be used as multi-value
descriptors of landscape patterns of a given area. This allows us to
calculate spatial signatures of two or more areas and then compare them
using a dissimilarity measure.

By default, the \texttt{window} argument is set to \texttt{NULL}, and
\textbf{motif} performs calculations for the entire area. However, the
package also offers the flexibility to compute numerous spatial
signatures for a single large area by dividing it into multiple subareas
referred to as ``local landscapes''. This partitioning is achieved by
specifying either a numeric value or an \texttt{sf} polygon data as the
\texttt{window} argument.

When a numeric value is provided, the input raster is subdivided into
square-shaped subareas, with each side length equal to the specified
numeric value, measured in cells. In the following example, we
demonstrate the calculation of co-occurrence vectors for square subareas
measuring 50 by 50 cells, which is equivalent to 5 by 5 kilometers in
this example.

\begin{Shaded}
\begin{Highlighting}[]
\NormalTok{france\_cove2 }\OtherTok{\textless{}{-}} \FunctionTok{lsp\_signature}\NormalTok{(france, }\AttributeTok{type=}\StringTok{"cove"}\NormalTok{, }\AttributeTok{window=}\DecValTok{50}\NormalTok{)}
\NormalTok{france\_cove2}
\end{Highlighting}
\end{Shaded}

\begin{verbatim}
# A tibble: 2,726 x 3
     id na_prop signature     
* <int>   <dbl> <list>        
1     1       0 <dbl [1 x 15]>
2     2       0 <dbl [1 x 15]>
3     3       0 <dbl [1 x 15]>
# i 2,723 more rows
\end{verbatim}

\begin{figure}

\centering{

\includegraphics[width=4.8in,height=\textheight]{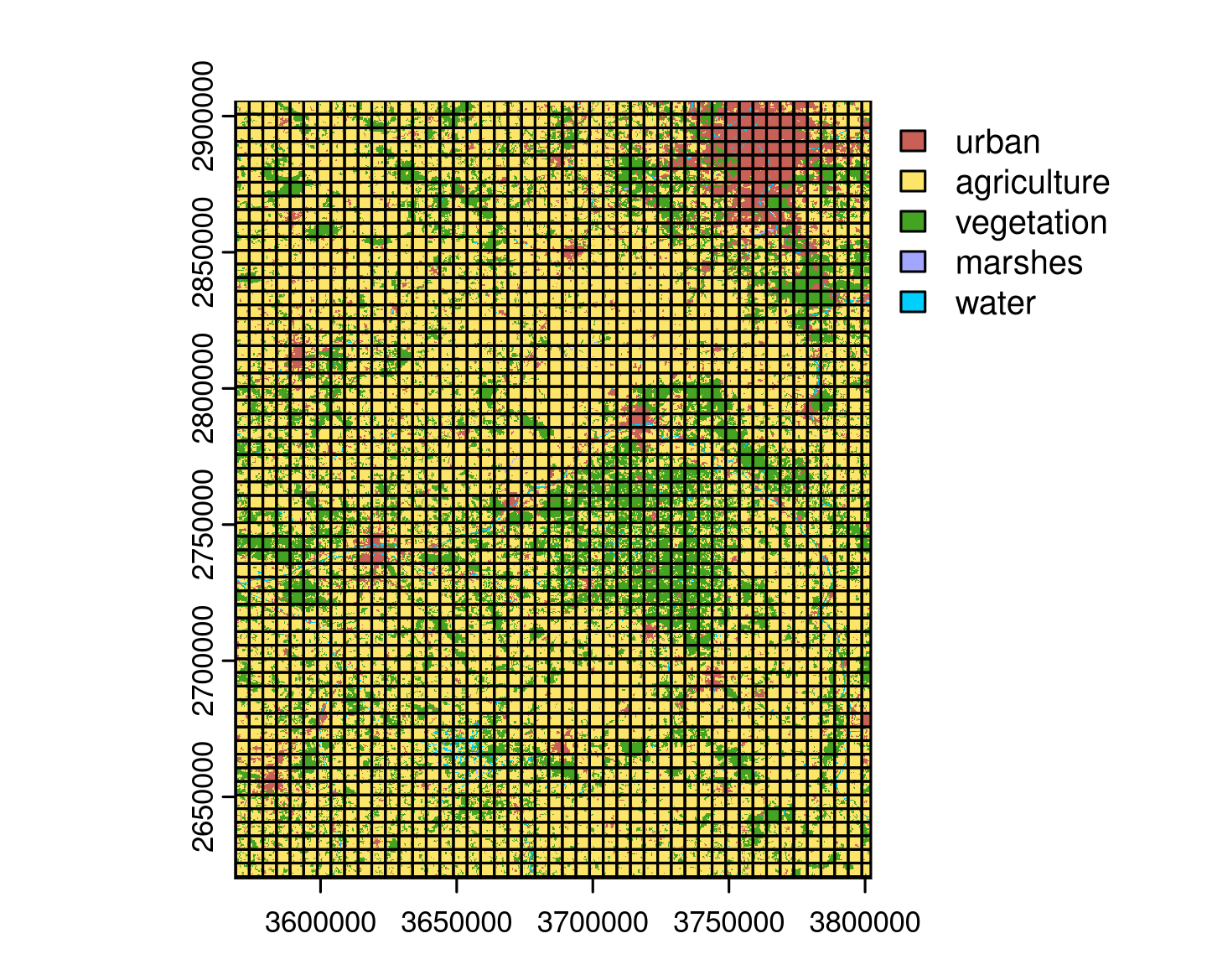}

}

\caption{\label{fig-grid}Land cover of the study area overlayed with a
grid with each subarea of 50 by 50 input raster cells}

\end{figure}

Now, notice that the output here is still a data frame with three
columns, but instead of having only one row, we now have 2,726 rows --
one row per subarea. We may visualize these subareas by creating a new
\texttt{sf} object with \texttt{lsp\_add\_sf()} and then using the
\texttt{plot()} function (Figure~\ref{fig-grid}).

\begin{Shaded}
\begin{Highlighting}[]
\NormalTok{france\_cove2\_sf }\OtherTok{\textless{}{-}} \FunctionTok{lsp\_add\_sf}\NormalTok{(france\_cove2)}
\FunctionTok{plot}\NormalTok{(france, }\AttributeTok{ext=}\FunctionTok{ext}\NormalTok{(france\_cove2\_sf))}
\FunctionTok{plot}\NormalTok{(}\FunctionTok{st\_geometry}\NormalTok{(france\_cove2\_sf), }\AttributeTok{add=}\ConstantTok{TRUE}\NormalTok{)}
\end{Highlighting}
\end{Shaded}

\begin{table}

\caption{\label{tbl-spatial-signatures}Spatial signatures available in
the motif package}

\centering{

\begin{tabular}[t]{l>{\raggedright\arraybackslash}p{3cm}>{\raggedright\arraybackslash}p{7cm}}
\toprule
type & name & description\\
\midrule
composition & Composition & A representation of the share of cells for each category within a local landscape. The length of the composition vector corresponds to the number of unique categories in a raster. Input: one categorical raster.\\
cove & Co-occurrence vector & A count of all of the pairs of the adjacent cells for each category in a local landscape. By default, its length equals to (n*n-n)/2+n, where n is a number of unique categories. Input: one categorical raster.\\
wecove & Weighted co-occurrence vector & A modification of a co-occurrence vector, in which each adjacency contributes to the output based on the values from the weight raster. By default, its length equals to (n*n-n)/2+n, where n is a number of unique categories. Input: one categorical and one numerical raster.\\
incove & Integrated co-occurrence vector & A count of all of the pairs of the adjacent cells for each category in a local landscape for all of the input rasters. Input: two or more categorical rasters.\\
- & A user-defined function & Any user-defined function that can summarize categorical spatial raster data.\\
\bottomrule
\end{tabular}

}

\end{table}%

The \texttt{window} argument can also accept an \texttt{sf} polygon,
allowing for the independent derivation of spatial signatures within
each polygon-defined area. In the provided example, we begin by randomly
selecting 100 sample points within the study area. For each point, we
create a buffer of 300 meters, resulting in a corresponding buffer
polygon. By setting this buffer polygon as the \texttt{window} argument,
we can obtain a spatial signature for the 300-meter buffer surrounding
each input point.

\begin{Shaded}
\begin{Highlighting}[]
\NormalTok{sample\_points }\OtherTok{\textless{}{-}} \FunctionTok{read\_sf}\NormalTok{(}\StringTok{"data/france\_sample\_points.gpkg"}\NormalTok{)}
\NormalTok{sample\_polys }\OtherTok{\textless{}{-}} \FunctionTok{st\_buffer}\NormalTok{(sample\_points, }\AttributeTok{dist=}\DecValTok{300}\NormalTok{)}
\NormalTok{france\_cove3 }\OtherTok{\textless{}{-}} \FunctionTok{lsp\_signature}\NormalTok{(france, }\AttributeTok{type=}\StringTok{"cove"}\NormalTok{, }\AttributeTok{window=}\NormalTok{sample\_polys)}
\NormalTok{france\_cove3 }\CommentTok{\# result not shown}
\end{Highlighting}
\end{Shaded}

\subsubsection{Landscape patterns'
comparison}\label{landscape-patterns-comparison}

Landscape patterns change through time and we are able to evaluate and
measure such a change. Here, we will use a second dataset, land cover
classes for the Centre-Val de Loire region in France for the year 2000.

\begin{Shaded}
\begin{Highlighting}[]
\NormalTok{france2000 }\OtherTok{\textless{}{-}} \FunctionTok{rast}\NormalTok{(}\StringTok{"data/raster\_france2000.tif"}\NormalTok{)}
\end{Highlighting}
\end{Shaded}

Now that we have two datasets, for the years 2000 and 2018, we can
proceed with their comparison. In pattern-based spatial analysis, the
comparison involves calculating the dissimilarity between the spatial
signatures derived from both rasters. This approach is complementary to
the pixel-based comparison that is based on comparing all the values for
each dataset, as it allows us to evaluate the change in the landscape
pattern.

To perform this comparison, we utilize the \texttt{lsp\_compare()}
function, and provide our two datasets, \texttt{france2000} and
\texttt{france}, along with specifying the type of spatial signature
(\texttt{type}) and the chosen dissimilarity measure
(\texttt{dist\_fun}). Additionally, we have the option to obtain the
results in one of several output classes. For instance, we can choose to
use the \texttt{SpatRaster} class from the \textbf{terra} package by
setting \texttt{output="terra"}.

\begin{Shaded}
\begin{Highlighting}[]
\NormalTok{lc\_change1 }\OtherTok{\textless{}{-}} \FunctionTok{lsp\_compare}\NormalTok{(france2000, france,}
                          \AttributeTok{type=}\StringTok{"cove"}\NormalTok{, }\AttributeTok{dist\_fun=}\StringTok{"jensen{-}shannon"}\NormalTok{, }
                          \AttributeTok{output=}\StringTok{"terra"}\NormalTok{)}
\NormalTok{lc\_change1[}\StringTok{"dist"}\NormalTok{]}
\end{Highlighting}
\end{Shaded}

\begin{verbatim}
class       : SpatRaster 
dimensions  : 1, 1, 1  (nrow, ncol, nlyr)
resolution  : 233300, 286000  (x, y)
extent      : 3568800, 3802100, 2619700, 2905700  (xmin, xmax, ymin, ymax)
coord. ref. : ETRS89-extended / LAEA Europe (EPSG:3035) 
source(s)   : memory
name        :         dist 
min value   : 0.0002682711 
max value   : 0.0002682711 
\end{verbatim}

\begin{figure}

\centering{

\includegraphics[width=4.8in,height=\textheight]{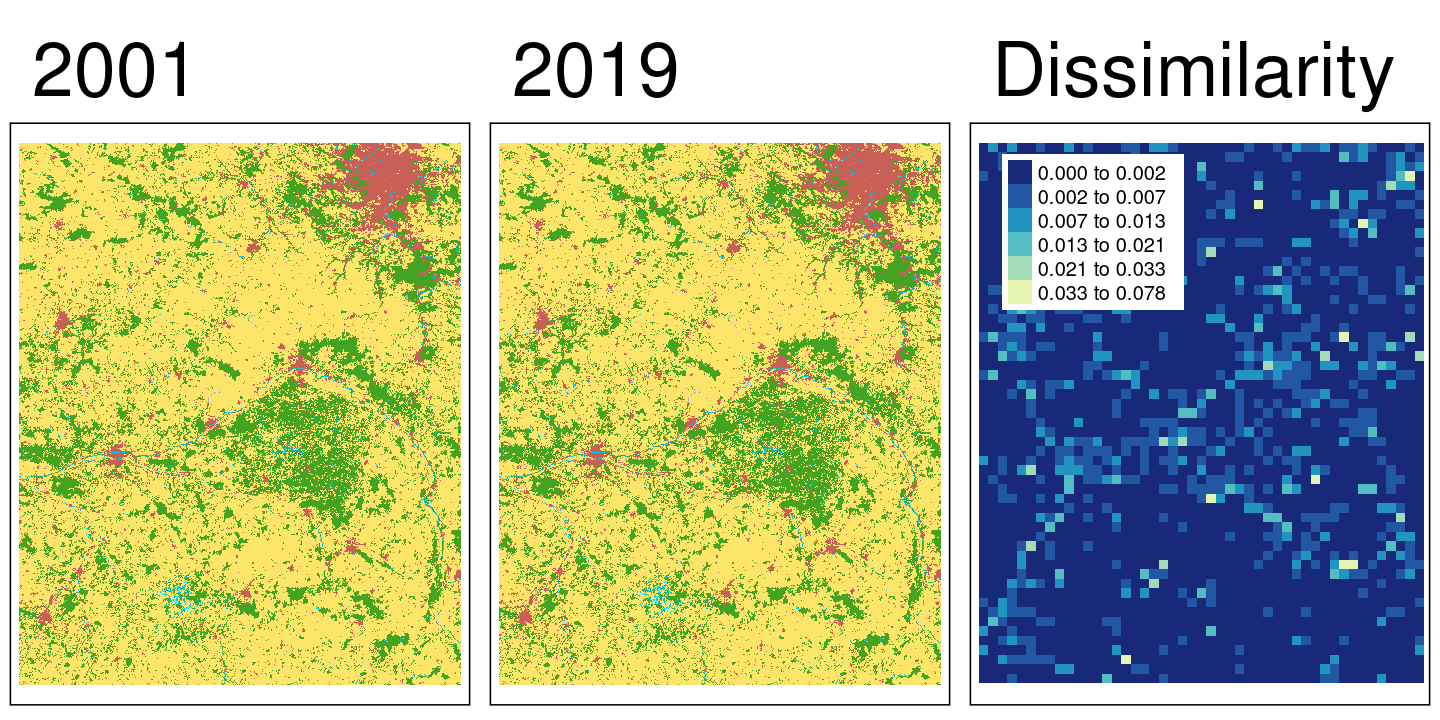}

}

\caption{\label{fig-change}(Left) land cover data for year 2000;
(center) land cover data for year 2018; (right) change in land cover
landscape patterns between 2000 and 2018 for 50 by 50 cells areas
(Jensen-Shannon distance)}

\end{figure}

The above example compared the change in landscape patterns using the
co-occurrence vector signature (\texttt{"cove"}) and the Jensen-Shannon
distance (\texttt{"jensen-shannon"}) for the whole area. However, we can
change the spatial scale of our analysis and look at the spatial pattern
changes also for smaller subareas. This can be done by adding a value to
the \texttt{window} parameter. In the example below, we set the
\texttt{window} parameter to 50, meaning each subarea will consist of 50
by 50 cells from the input rasters Figure~\ref{fig-change}.

\begin{Shaded}
\begin{Highlighting}[]
\NormalTok{lc\_change }\OtherTok{\textless{}{-}} \FunctionTok{lsp\_compare}\NormalTok{(france2000, france,}
                         \AttributeTok{type=}\StringTok{"cove"}\NormalTok{, }\AttributeTok{dist\_fun=}\StringTok{"jensen{-}shannon"}\NormalTok{, }
                         \AttributeTok{output=}\StringTok{"terra"}\NormalTok{, }\AttributeTok{window=}\DecValTok{50}\NormalTok{)}
\end{Highlighting}
\end{Shaded}

The highest values in the right panel of Figure~\ref{fig-change}
represent areas with the most prominent change of land cover landscape
patterns. They mainly relate to the change of herbaceous areas into
shrubs. We can also find examples of the areas with the largest changes
by subsetting only the local landscapes with dissimilarity above some
threshold. Then, \texttt{the\ lsp\_extract()} function can be used to
extract a local landscape with a given \texttt{id}.

\begin{Shaded}
\begin{Highlighting}[]
\NormalTok{lc\_change\_df }\OtherTok{\textless{}{-}} \FunctionTok{as.data.frame}\NormalTok{(lc\_change)}
\FunctionTok{subset}\NormalTok{(lc\_change\_df, dist}\SpecialCharTok{\textgreater{}}\FloatTok{0.05}\NormalTok{)}
\end{Highlighting}
\end{Shaded}

\begin{verbatim}
       id na_prop_x na_prop_y       dist
1764 1764         0         0 0.05966520
2107 2107         0         0 0.07848007
\end{verbatim}

\begin{Shaded}
\begin{Highlighting}[]
\NormalTok{compare\_1 }\OtherTok{\textless{}{-}} \FunctionTok{lsp\_extract}\NormalTok{(}\FunctionTok{c}\NormalTok{(france2000, france), }\AttributeTok{window=}\DecValTok{50}\NormalTok{, }\AttributeTok{id=}\DecValTok{2107}\NormalTok{)}
\FunctionTok{plot}\NormalTok{(compare\_1, }\AttributeTok{legend =} \ConstantTok{FALSE}\NormalTok{)}
\end{Highlighting}
\end{Shaded}

\begin{figure}

\centering{

\includegraphics[width=4.8in,height=\textheight]{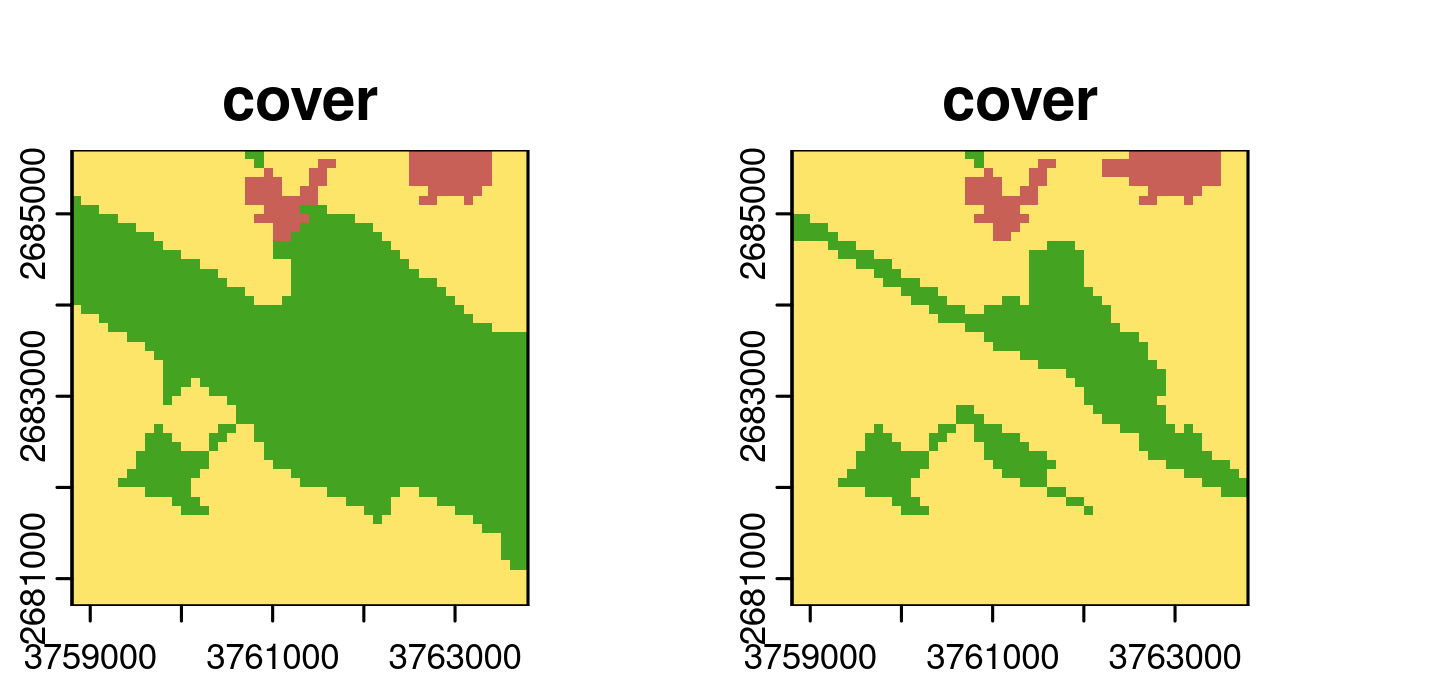}

}

\caption{\label{fig-change2}An example of a local landscape of 5 by 5
kilometers with the largest change of land cover landscape patterns
between 2000 and 2018}

\end{figure}%

Figure~\ref{fig-change2} shows a local landscape with the largest change
of land cover landscape patterns between 2000 and 2018. This location is
an example of an area with a large loss of natural vegetation (to
agriculture) and a small expansion of urban areas.

\subsubsection{Landscape patterns'
search}\label{landscape-patterns-search}

Another example of pattern-based spatial analysis is searching for areas
with similar landscape patterns to a query one. Pattern-based search is
a comparison of a spatial pattern of a given local landscape to many
subareas of a larger raster.

Here, we start by reading polygon data with our study area and cropping
the \texttt{france} raster to its borders. Thus, the
\texttt{france\_study\_area} object is a raster storing LULC data for
the study area.

\begin{Shaded}
\begin{Highlighting}[]
\NormalTok{study\_area }\OtherTok{\textless{}{-}} \FunctionTok{read\_sf}\NormalTok{(}\StringTok{"data/france\_study\_area.gpkg"}\NormalTok{)}
\NormalTok{france\_study\_area }\OtherTok{\textless{}{-}} \FunctionTok{crop}\NormalTok{(france, study\_area, }\AttributeTok{mask=}\ConstantTok{TRUE}\NormalTok{)}
\end{Highlighting}
\end{Shaded}

Search can be performed with the \texttt{lsp\_search()} function, while
providing two raster datasets (an area of interest and a larger area),
the signature type, distance measure and the window size.

\begin{Shaded}
\begin{Highlighting}[]
\NormalTok{nlcd\_search }\OtherTok{\textless{}{-}} \FunctionTok{lsp\_search}\NormalTok{(france\_study\_area, france,}
                         \AttributeTok{type=}\StringTok{"cove"}\NormalTok{, }\AttributeTok{dist\_fun=}\StringTok{"jensen{-}shannon"}\NormalTok{,}
                         \AttributeTok{output=}\StringTok{"terra"}\NormalTok{, }\AttributeTok{window=}\DecValTok{50}\NormalTok{)}
\end{Highlighting}
\end{Shaded}

\begin{Shaded}
\begin{Highlighting}[]
\CommentTok{\# result not shown}
\FunctionTok{plot}\NormalTok{(nlcd\_search[[}\StringTok{"dist"}\NormalTok{]])}
\end{Highlighting}
\end{Shaded}

\begin{figure}

\centering{

\includegraphics[width=4.8in,height=\textheight]{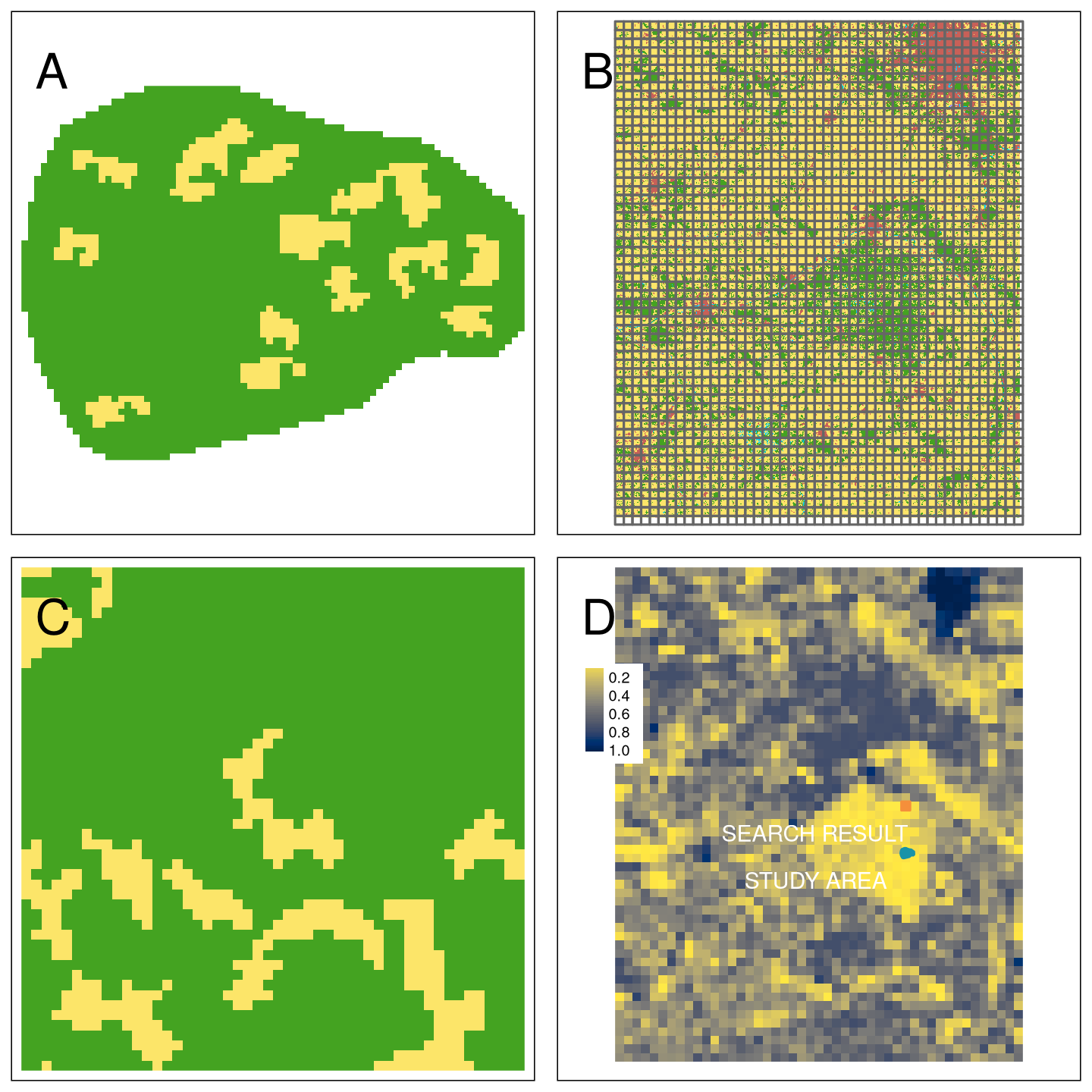}

}

\caption{\label{fig-search}(A) Area of interest, (B) Search area divided
into a set of regular local landscapes, (C) A local landscape with the
most similar land cover spatial pattern to the area of interest, (D)
Dissimilarity between land cover landscape patterns of the area of
interest and the search area}

\end{figure}

Its output is a \textbf{terra} object with three raster layers: a unique
\texttt{id}, the proportion of cells in an area with missing values
(\texttt{na\_prop}), and the calculated dissimilarity (\texttt{dist})
(Figure~\ref{fig-search} D). The areas with the smallest dissimilarity
values are the ones that have the most similar spatial pattern to the
area of interest.

\begin{Shaded}
\begin{Highlighting}[]
\NormalTok{nlcd\_search\_df }\OtherTok{\textless{}{-}} \FunctionTok{as.data.frame}\NormalTok{(nlcd\_search)}
\FunctionTok{subset}\NormalTok{(nlcd\_search\_df, dist}\SpecialCharTok{\textless{}}\FloatTok{0.001}\NormalTok{)}
\end{Highlighting}
\end{Shaded}

\begin{verbatim}
       id na_prop         dist
1303 1303       0 0.0003939899
1677 1677       0 0.0007520309
1867 1867       0 0.0007956902
\end{verbatim}

To extract a selected local landscape, the \texttt{lsp\_extract()}
function can also be used here (Figure~\ref{fig-search} C).

\begin{Shaded}
\begin{Highlighting}[]
\NormalTok{search\_1 }\OtherTok{\textless{}{-}} \FunctionTok{lsp\_extract}\NormalTok{(france, }\AttributeTok{window=}\DecValTok{50}\NormalTok{, }\AttributeTok{id=}\DecValTok{1303}\NormalTok{)}
\end{Highlighting}
\end{Shaded}

\subsubsection{Landscape patterns' regionalization and
clustering}\label{landscape-patterns-regionalization-and-clustering}

Spatial structures derived with \texttt{lsp\_signature()} may also be
converted to spatial objects of \texttt{sf}, \texttt{stars} and
\texttt{terra} classes. The \texttt{lsp\_add\_terra()} function, for
example, takes a spatial signature object and converts in into a
\textbf{terra} raster object.

\begin{Shaded}
\begin{Highlighting}[]
\NormalTok{france\_cove2 }\OtherTok{\textless{}{-}} \FunctionTok{lsp\_signature}\NormalTok{(france, }\AttributeTok{type=}\StringTok{"cove"}\NormalTok{, }\AttributeTok{window=}\DecValTok{10}\NormalTok{)}
\NormalTok{france\_cove\_terra }\OtherTok{\textless{}{-}} \FunctionTok{lsp\_add\_terra}\NormalTok{(france\_cove2, }\AttributeTok{metadata=}\ConstantTok{FALSE}\NormalTok{)}
\NormalTok{france\_cove\_terra}
\end{Highlighting}
\end{Shaded}

\begin{verbatim}
class       : SpatRaster 
dimensions  : 286, 234, 15  (nrow, ncol, nlyr)
resolution  : 1000, 1000  (x, y)
extent      : 3568800, 3802800, 2619700, 2905700  (xmin, xmax, ymin, ymax)
coord. ref. : ETRS89-extended / LAEA Europe (EPSG:3035) 
source(s)   : memory
names       : X1,        X2, X3,        X4,        X5, X6, ... 
min values  :  0, 0.0000000,  0, 0.0000000, 0.0000000,  0, ... 
max values  :  1, 0.3191489,  1, 0.3055556, 0.3944444,  1, ... 
\end{verbatim}

This new object has several raster layers representing spatial signature
elements. This opens new possibilities for using the information of
landscape patterns in other workflows related to spatial
regionalization, clustering, and machine learning.

A regionalization example below uses the \textbf{supercells} package to
group adjacent local areas with similar landscape patterns. We can apply
the \textbf{supercells} algorithm (Nowosad and Stepinski 2022), which
expects four essential arguments: input raster data, \texttt{k}: a
number of regions desired by the user, \texttt{compactness}: a
compactness value, where larger values cause clusters to be more
compact, and \texttt{dist\_fun}: a distance function.

\begin{Shaded}
\begin{Highlighting}[]
\FunctionTok{library}\NormalTok{(supercells)}
\NormalTok{france\_sc }\OtherTok{\textless{}{-}} \FunctionTok{supercells}\NormalTok{(france\_cove\_terra, }\AttributeTok{k=}\DecValTok{200}\NormalTok{, }\AttributeTok{compactness=}\FloatTok{0.6}\NormalTok{,}
                        \AttributeTok{dist\_fun=}\StringTok{"jensen{-}shannon"}\NormalTok{, }\AttributeTok{metadata=}\ConstantTok{FALSE}\NormalTok{)}
\end{Highlighting}
\end{Shaded}

\begin{figure}

\centering{

\includegraphics[width=4.8in,height=\textheight]{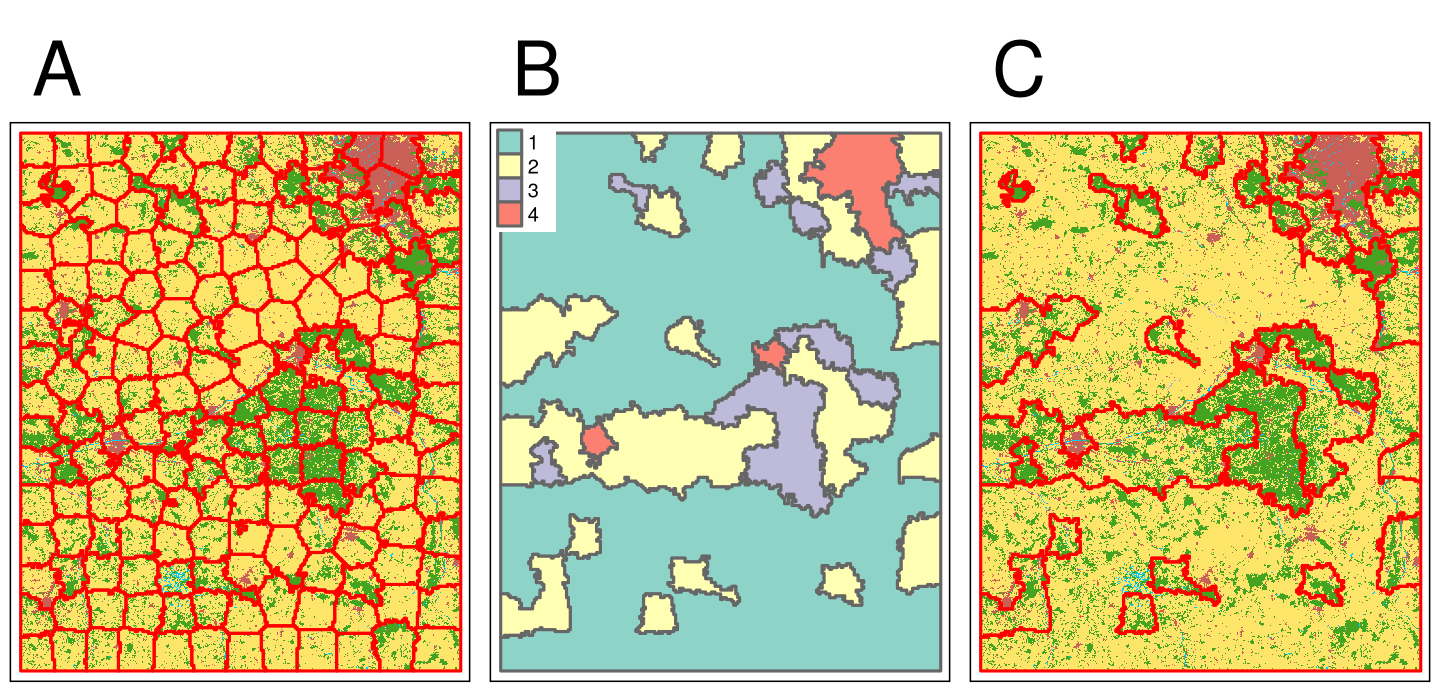}

}

\caption{\label{fig-supercells}(A) Study area divided into supercells of
homogeneous land cover landscape patterns, (B) Supercells grouped into
four clusters using the hierarchical clustering method, (C) Borders of
the polygons derived with the hierarchical clustering method}

\end{figure}

The output here is a spatial vector with a set of polygons, where each
polygon represents a larger area with a certain level of homogeneity of
its landscape patterns (Figure~\ref{fig-supercells} A).

\begin{Shaded}
\begin{Highlighting}[]
\FunctionTok{plot}\NormalTok{(france)}
\FunctionTok{plot}\NormalTok{(}\FunctionTok{st\_geometry}\NormalTok{(france\_sc), }\AttributeTok{add=}\ConstantTok{TRUE}\NormalTok{, }\AttributeTok{border=}\StringTok{"red"}\NormalTok{) }\CommentTok{\# result not shown}
\end{Highlighting}
\end{Shaded}

You may notice that, while the land cover landscape patterns are
homogeneous internally, there are several examples of similar adjacent
polygons. In cases like this, we could merge such polygons, for example,
using a clustering method, such as hierarchical clustering or k-means.
Firstly, we need to drop the geometry column from the
\texttt{france\_sc} object, as it is not needed for the clustering, and
then calculate the value distances between the landscape patterns of
each supercell. Next, we can apply the hierarchical clustering method
with the \texttt{hclust()} function. Based on the dendrogram
visualization (\texttt{plot(hc)}), we can select the number of clusters:
here we choose four clusters. Finally, we can assign a cluster number to
each supercell using the \texttt{cutree()} function.

\begin{Shaded}
\begin{Highlighting}[]
\NormalTok{france\_sc\_df }\OtherTok{\textless{}{-}} \FunctionTok{st\_drop\_geometry}\NormalTok{(france\_sc) }
\NormalTok{france\_sc\_dist }\OtherTok{\textless{}{-}}\NormalTok{ philentropy}\SpecialCharTok{::}\FunctionTok{distance}\NormalTok{(france\_sc\_df, }
                                        \AttributeTok{method=}\StringTok{"jensen{-}shannon"}\NormalTok{,}
                                        \AttributeTok{as.dist.obj=}\ConstantTok{TRUE}\NormalTok{)}
\NormalTok{hc }\OtherTok{\textless{}{-}} \FunctionTok{hclust}\NormalTok{(france\_sc\_dist)}
\NormalTok{france\_sc}\SpecialCharTok{$}\NormalTok{k }\OtherTok{\textless{}{-}} \FunctionTok{cutree}\NormalTok{(hc, }\DecValTok{4}\NormalTok{)}
\end{Highlighting}
\end{Shaded}

The hierarchical clustering method assigns a cluster number to each,
previously derived supercell. Thus, to obtain a separate polygon for
each cluster, we need to perform a postprocessing step:
\texttt{aggregate()} the clusters to dissolve the borders between
clusters.

\begin{Shaded}
\begin{Highlighting}[]
\NormalTok{france\_sc2 }\OtherTok{\textless{}{-}} \FunctionTok{aggregate}\NormalTok{(france\_sc, }\AttributeTok{by=}\FunctionTok{list}\NormalTok{(france\_sc}\SpecialCharTok{$}\NormalTok{k), }\AttributeTok{FUN=}\NormalTok{mean)}
\end{Highlighting}
\end{Shaded}

The dissolved areas of clusters are shown on panel B of
Figure~\ref{fig-supercells}, while the C panel shows the final created
polygons. The first cluster represents mostly agricultural areas, the
third cluster represents mostly forested areas, and the fourth cluster
represents mostly urban areas. The second cluster is a bit more complex,
as it contains a mosaic of forested and agricultural areas. The result
seems mostly satisfactory, where most created polygons are internally
homogeneous and visibly distinct from their neighbors. At the same time,
there are a few areas in which homogeneity needs to be improved. This
suggests that the result could still be improved, for example, by using
a larger number of initial supercells or changing the number of expected
clusters.

\subsection{Conclusions}\label{conclusions}

In conclusion, this chapter has highlighted the significance of
categorical raster data, specifically land use or land cover (LULC)
data, in ecological studies. We have explored the role of landscape
metrics and pattern-based spatial analysis in quantifying and analyzing
landscape patterns, using the R packages \textbf{landscapemetrics} and
\textbf{motif}.

Landscape metrics, computed using the \textbf{landscapemetrics} package,
have proven to be valuable tools in spatial ecology and landscape
ecology. They allow us to quantify the composition and configuration of
spatial landscape characteristics, enabling a deeper understanding of
environmental processes and changes. By calculating metrics at various
levels and visualizing them, researchers can gain insights into the
patterns and structures present in the landscape. However, it is crucial
to select metrics purposefully, aligned with research questions and
hypotheses, to avoid the pitfall of ``metric fishing'' (Gustafson 2019).

Moreover, pattern-based spatial analysis, facilitated by the
\textbf{motif} package, offers a robust framework for analyzing and
comparing landscape patterns. By considering local groups of adjacent
cells as landscape patterns and calculating spatial signatures, we can
capture both composition and configuration information. This approach
allows us to compare landscape patterns over time, identify areas with
similar patterns, and group similar areas together. The \textbf{motif}
package's ability to handle large raster datasets is particularly
advantageous.

The chapter has provided practical examples and code snippets
demonstrating the calculation of landscape metrics and spatial
signatures, as well as their visualization. We have shown how to use
these techniques to analyze landscape patterns, explore changes over
time, and extract local landscapes with significant pattern variations.
By combining the power of landscape metrics and pattern-based spatial
analysis, researchers can gain a comprehensive understanding of
landscape dynamics, composition, and configuration. These techniques
facilitate the investigation of the pattern-process link that is a
central framework of many ecological fields, but specifically landscape
ecology (Turner 1989). For example, the pattern-process link using LULC
data was studied for species distributions (Marshall et al. 2018),
(functional) connectivity (Vogt et al. 2009), or genetic population
structure (Borthwick et al. 2020). However, patterns and processes are
often in an interacting relationship influencing each other, and many
drivers and processes act simultaneously - such as climate,
disturbances, succession, biotic competition and dispersal, or human
land use (Turner 2005). The pattern-process link is a complex and
challenging topic, and the analysis of landscape patterns is only one
part of the puzzle. Nevertheless, landscape metrics and pattern-based
spatial analysis are powerful tools to investigate the pattern-process
link and to gain a better understanding of the landscape.

Both packages, being open-source, offer extensive possibilities for
customization and expansion. This enables researchers to tailor them to
their specific inquiries and requirements. The \textbf{landscapemetrics}
package, for instance, allows for the integration of new metrics through
the utilization of diverse utility functions, thereby enhancing its
capabilities. Similarly, the \textbf{motif} package can be expanded by
incorporating user-defined spatial signatures, which can be used to
measure similarities between different properties of landscape patterns.
Consequently, contributions to both packages are highly encouraged and
welcomed. You can actively participate in the development of these
packages at \url{https://github.com/r-spatialecology/landscapemetrics}
and \url{https://github.com/Nowosad/motif}.

\subsubsection{Acknowledgments}\label{acknowledgments}

This work was supported by the Initiative of Excellence - Research
University project at Adam Mickiewicz University, Poznan {[}grant number
107/07/POB1/0002{]}. We also thank the developers and contributors of
the \textbf{landscapemetrics} and \textbf{motif} packages for their work
and support.

\subsubsection*{References}\label{references}
\addcontentsline{toc}{subsubsection}{References}

\phantomsection\label{refs}

\end{document}